\documentclass[5p,a4paper]{elsarticle}
\makeatletter
\def\ps@pprintTitle{%
 \let\@oddhead\@empty
 \let\@evenhead\@empty
 \def\@oddfoot{\centerline{\thepage}}%
 \let\@evenfoot\@oddfoot}
\makeatother
\usepackage{amssymb}
\usepackage{color}
\usepackage{bm}
\usepackage{textcomp}

\definecolor{orange}{rgb}{1,0.5,0}
\newcommand{\figref}[1]{Figure~\ref{#1}}

\renewcommand{\vec}[1]{\mathbf{#1}}
\newcommand{\punc}[1]{\,#1}

\newcommand{\tabref}[1]{Table~(\ref{#1})}
\newcommand{\degreeC}{$^{\circ}\!$C}
\newcommand{\pme}[2]{$#1\!\pm\!#2$}

\biboptions{sort&compress}

\begin{document}

\title{Design of a nickel-base superalloy using a neural~network}
\author{B.D.~Conduit}
\address{Rolls-Royce plc, PO Box 31, Derby, DE24 8BJ, United Kingdom}
\author{N.G.~Jones}
\address{Rolls-Royce UTC, 27 Charles Babbage Road, Cambridge, CB3 0FS, United Kingdom}
\author{H.J.~Stone}
\address{Rolls-Royce UTC, 27 Charles Babbage Road, Cambridge, CB3 0FS, United Kingdom}
\author{G.J.~Conduit}
\address{Cavendish Laboratory, J.J. Thomson Avenue, Cambridge, CB3 0HE, United Kingdom}
\date{\today}


\begin{abstract}
  A new computational tool has been developed to model, discover, and
  optimize new alloys that simultaneously satisfy up to eleven physical
  criteria. An artificial neural network is trained from pre-existing
  materials data that enables the prediction of individual material
  properties both as a function of composition and heat treatment routine,
  which allows it to optimize the material properties to search for the
  material with properties most likely to exceed a target criteria. We
  design a new polycrystalline nickel-base superalloy with the optimal
  combination of cost, density, $\gamma'$ phase content and solvus, phase
  stability, fatigue life, yield stress, ultimate tensile strength, stress
  rupture, oxidation resistance, and tensile elongation. Experimental data
  demonstrates that the proposed alloy fulfills the computational
  predictions, possessing multiple physical properties, particularly
  oxidation resistance and yield stress, that exceed existing commercially
  available alloys.
\end{abstract}

\maketitle

Despite the central importance of materials in enabling new technologies,
historically the only way to develop new materials has been through
experiment driven trial and improvement~\cite{Curtarolo13}. This means that
commercially available alloys are the result of many years of empirical
development, and whilst they have good properties, they do not necessarily
offer the right balance of properties needed for specific engineering
applications. The capability to discover materials computationally has the
potential to empower engineers to utilize materials optimized for their
application~\cite{Kuehmann09}. The development of new algorithms and a surge
in computing power has enabled the screening of large numbers of prospective
compositions with first principles calculations~\cite{Curtarolo13}.
Designing alloy compositions to identify which best fulfill the target
criteria has previously been attempted with a Pareto
set~\cite{Bligaard03,Greeley06,Lejaeghere13}, a principal component
analysis~\cite{Toda13}, robust design~\cite{Backman06}, and the orthogonal
optimization of different
properties~\cite{Joo09,Xu09,Reed09,Kuehmann09,Tancret13}. In this paper, we
develop a new computational tool that combines experimental data with
computational thermodynamic predictions~\cite{thermocalcgeneral02} to
rapidly, reliably, and robustly identify the alloy composition that is most
likely to meet a multi-criterion specification~\cite{Conduit2014iv}.

We use the tool to propose a new nickel-base
superalloy~\cite{Conduit2014vii}. A nickel-base superalloy is an ideal case
study for real-life materials design, because of the need to obtain the
optimal balance of many properties, including physical and thermodynamic
requirements, with a special focus on improving the critical properties of
yield stress and oxidation resistance.  This case study not only serves as
an independent test of the alloy design approach, but moreover leads to a
potentially commercially viable alloy.

In the first part of this paper, the computational tools developed to
predict both the expected value and the associated uncertainty in a
material's physical properties are described. The computational tools are
used to evaluate the likelihood that a proposed alloy composition will
satisfy the design criteria, and to then select the composition most likely
to fulfill the design criteria. This means that we are proposing the alloy
that is most likely to succeed in experimental verification, and, therefore,
be of most use to the engineer. The efficiency of this approach is
demonstrated in the second section of the paper where we present
experimental results for the properties of the proposed new alloy, proving
that is has a combination of properties that surpass commercially available
alternatives.

\section{Formalism}

The goal of the concurrent materials design formalism is to predict a
composition and processing variables that are most likely to fulfill the
multi-criteria target specification such as a maximum permissible cost and
minimum allowed yield stress. First, predictive models are constructed for
each property, second, these models are used to calculate the probability
that a proposed composition fulfills the target specification, and finally
search composition space for the alloy most likely to fulfill the
specification.

\begin{table}
 \small\centering
 \begin{tabular}{lccc}
   Property&Approach&Points&Target\\
   \hline\hline
   Cost&Physical&\cite{AmericanElements2013}&$<$33.7kg$^{-1}$\\
   Density&Physical&\cite{reed2009}&$<$8281kgm$^{-3}$\\
   $\gamma'$ content&{\sc calphad}&\cite{thermocalcgeneral02,connor2009}&$<$50.4vol$\%$\\
   \hline
   Stability&{\sc calphad}&\cite{thermocalcgeneral02}&$>$99.0vol$\%$\\
   Fatigue\,life&Neural\,net&15105\cite{specialmetals2013,haynesinternational2013}&$>10^{3.9}$cycles\\
   Yield stress&Neural\,net&6939\cite{tomasello1996,radavich2004,sims1987,ewing1976,sato1993,mannan2000,mitchell2004,meng1984,tien1989,huron2004,sharma1983,cowen2008,pike2008,mannan2004,tillack1991,rizzo1991,moll1971,brinegar1984,bouse1989,braun1989,chang1989,jackman1991,schirra1991,guo1991,mannan2003,rizzo1968,radavich1984,xie1996,loewenkamp1988,quested1988,seib2000,shaw1980,sjoeberg2004,nganbe2009,kaufman1984,eng1980,brinegar1984_2,specialmetals1971,locq2000,barker1972,hunt2001,wanner1992,tien1990,gu2008,couturier2004,sczerzenie1984,jain2000,green1996,furrer2000,ferrari1976,raisson1976,richards1968,maurer1980}&$>$752.2MPa\\
   UTS&Neural\,net&
   6693\cite{tomasello1996,radavich2004,sims1987,ewing1976,sato1993,mannan2000,mitchell2004,meng1984,tien1989,huron2004,sharma1983,cowen2008,pike2008,mannan2004,tillack1991,rizzo1991,moll1971,brinegar1984,bouse1989,braun1989,chang1989,jackman1991,schirra1991,guo1991,mannan2003,rizzo1968,radavich1984,xie1996,loewenkamp1988,quested1988,seib2000,shaw1980,sjoeberg2004,nganbe2009,kaufman1984,eng1980,brinegar1984_2,specialmetals1971,locq2000,barker1972,hunt2001,wanner1992,tien1990,gu2008,couturier2004,sczerzenie1984,jain2000,green1996,furrer2000,ferrari1976,raisson1976,richards1968,maurer1980}&$>$960.0MPa\\
   300hr\,rupture&Neural\,net&10860\cite{tomasello1996,radavich2004,sims1987,ewing1976,sato1993,mannan2000,mitchell2004,meng1984,tien1989,huron2004,sharma1983,cowen2008,pike2008,mannan2004,tillack1991,rizzo1991,moll1971,brinegar1984,bouse1989,braun1989,chang1989,jackman1991,schirra1991,guo1991,mannan2003,rizzo1968,radavich1984,xie1996,loewenkamp1988,quested1988,seib2000,shaw1980,sjoeberg2004,nganbe2009,kaufman1984,eng1980,brinegar1984_2,specialmetals1971,locq2000,barker1972,hunt2001,wanner1992,tien1990,gu2008,couturier2004,sczerzenie1984,jain2000,green1996,furrer2000,ferrari1976,raisson1976,richards1968,maurer1980}&$>$674.5MPa\\
   Cr\,activity&Neural\,net&915\cite{reed2010}&$>$0.14\\
   $\gamma$' solvus&{\sc calphad}&\cite{thermocalcgeneral02}&$>$983$^{\circ}\rm{C}$\\
   Elongation&Neural\,net&2248\cite{tomasello1996,radavich2004,sims1987,ewing1976,sato1993,mannan2000,mitchell2004,meng1984,tien1989,huron2004,sharma1983,cowen2008,pike2008,mannan2004,tillack1991,rizzo1991,moll1971,brinegar1984,bouse1989,braun1989,chang1989,jackman1991,schirra1991,guo1991,mannan2003,rizzo1968,radavich1984,xie1996,loewenkamp1988,quested1988,seib2000,shaw1980,sjoeberg2004,nganbe2009,kaufman1984,eng1980,brinegar1984_2,specialmetals1971,locq2000,barker1972,hunt2001,wanner1992,tien1990,gu2008,couturier2004,sczerzenie1984,jain2000,green1996,furrer2000,ferrari1976,raisson1976,richards1968,maurer1980}&$>$11.6$\%$\\
 \end{tabular}
 \caption{
   The approach used to predict
   properties, number of experimental points used to train the nickel-base superalloy neural network
   models, references for the data, and the target specification.}
 \label{tbl:DataSources}
\end{table}

The properties that were optimized in the design of the nickel-base
superalloy are shown in \tabref{tbl:DataSources}. With properties depending
on contrasting length scales, different calculation methods were adopted
that are referenced in the tables. For some properties including cost and
density, physically based models were adopted: density was calculated as a
weighted average of the densities of the elements comprising the alloy, cost
was a weighted average of the prices of the elements comprising the alloy
plus a fixed charge for the preparation and heat treatments. Two
  methods can be used to calculate phase stability, the first using the
  PhaComp method~\cite{Morinaga84} is evaluated from the average
  energy levels of d-orbitals ($M_{\rm{d}}$) within the transition metals
  in the alloy, a $M_{\rm{d}}<0.98$eV indicates that there will be
  acceptably low formation of topologically closed packed
  phases. $M_{\rm{d}}$ can be evaluated with little computational expense
  for the composition space search; the final composition was cross-checked
with the CALPHAD method~\cite{Kaufman70}, with data sourced from the TTNI8
database~\cite{thermocalcgeneral02}.

\begin{figure}
 \centering
 \includegraphics[width=0.7\linewidth]{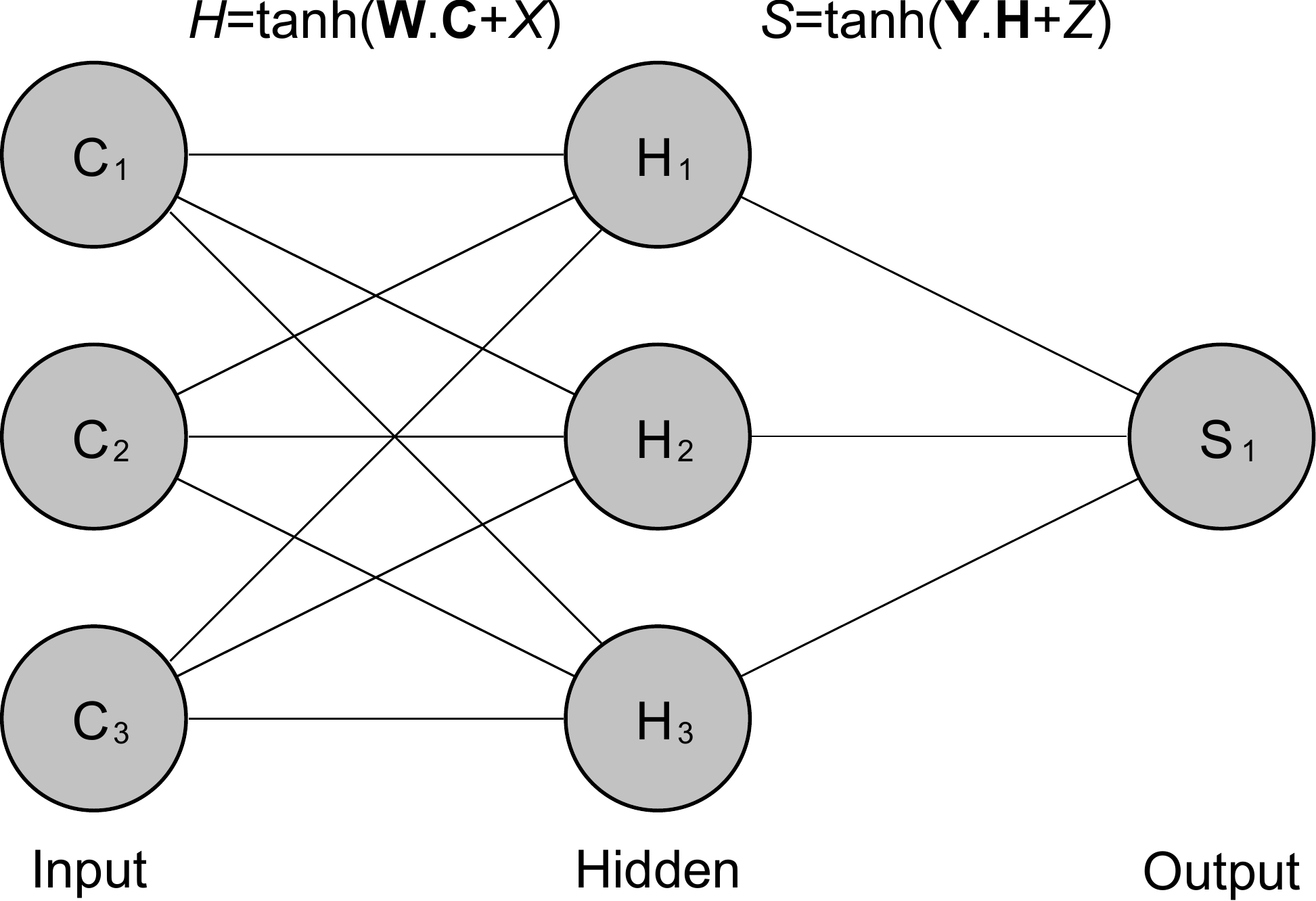}
 \caption{(Color online) Schematic representation of the neural network.}
 \label{fig:NeuralNetwork}
\end{figure}

Several properties cannot be reliably calculated from computer
  modeling. Instead, for the fatigue life, yield stress, ultimate tensile
  strength, rupture, and tensile elongation a database of experimental
  results as a function of composition and heat treatment was compiled from
  the sources referenced in \tabref{tbl:DataSources}.  A neural network
model was constructed that predicts the physical properties,
$\vec{S}(\vec{C})$, for composition, $\vec{C}$. The form of the neural
network is shown in \figref{fig:NeuralNetwork}.  The network takes three
input variables $C_{1,2,3}$, the links act on the variables through the
indicator function $H_{1,2,3}=\tanh(\vec{W_{1,2,3}}\cdot\vec{C}+X_{1,2,3})$,
to transform them into three hidden node values $H_{1,2,3}$. The hidden
nodes are again combined by indicator functions
$\tanh(\vec{Y}\cdot\vec{H}+Z)$ to give the final output value $S_{1}$. This
network, shown in \figref{fig:NeuralNetwork}, has free variables
$W_{11,12,13,21,22,23,31,32,33}$, $X_{1,2,3}$, $Y_{1,2,3}$ and $Z$. In the
design of the nickel-base superalloy there are up to $N_{\rm{D}}=25$ input
values $C_{1,2,\cdots,25}$ covering both composition and heat
treatments. The free variables are optimized by minimizing the reduced
chi-squared statistic over the preexisting data
\begin{equation}
 \chi_{\rm{red}}^{2}=\frac{1}{N-(2+N_{\rm{D}})N_{\rm{H}}-1}
 \sum_{j=1}^{N}\frac{(S_{j}-s_{j})^2}{\sigma_{j}^2}\punc{.}
\end{equation}
where $N$ is the number of data points of values $\{s\}$ and experimental
uncertainty $\{\sigma\}$ available for training.  The statistic divides by
the number of free variables $N-(2+N_{\rm{D}})N_{\rm{H}}-1$ that
includes the number of hidden nodes $N_{\rm{H}}$. We found that typically
$N_{\rm{H}}=3$ hidden nodes gave the minimal reduced chi-squared
statistic.  A separate system of neural networks with its own optimizable
parameters was constructed for each output $\vec{S}_{i}$.  To evaluate the
uncertainty in the predictions, a committee of $M=64$ neural network models,
labeled $j\in\{1\dots M\}$, was constructed using the Bayesian bootstrap
approach~\cite{Heskes97,Papadopoulos01} that delivers predictions
$\vec{S}(\vec{C})_{j}$. Each neural network model was constructed by
associating random weights with the input data, which deliver a range of
outputs correctly distributed to reflect the underlying uncertainty in the
networks due to the limited input data~\cite{Heskes97,Papadopoulos01}.  For
the proposed composition, the average value over the models gives the
predicted physical value
$\vec{V}_{\!\vec{C}}=\frac{1}{M}\sum_{j=1}^{M}\vec{S}_{\vec{C},j}$, and the
uncertainty was found through the covariance matrix
$\bm{\Sigma}_{\vec{C}}=\sum_{j=1}^{M}
(\vec{S}_{\vec{C},j}-\vec{V}_{\!\vec{C}})
(\vec{S}_{\vec{C},j}-\vec{V}_{\!\vec{C}})^{\rm{T}}$~\cite{Wasserman04}.
The knowledge of uncertainty is crucial as it allows the designer to balance
the risk of materials with lower uncertainty but are less capable versus
promising but speculative alloys.

With a prescription to calculate the separate properties of a material, a
single merit index is defined to describe how well the material satisfies
the design criteria that can then be optimized. The neural network models
offer a unique insight into the inevitable uncertainty that exists in the
predictions based on experimental data. The uncertainty means that the
probability that a putative composition will satisfy the target design
criterion, $\vec{T}$, is
$P_{\vec{C}}=\Phi[\bm{\Sigma}_{\vec{C}}^{-1}(\vec{V}_{\!\vec{C}}-\vec{T})]$,
where we assume that uncertainties are normally distributed so $\Phi$ is the
multivariate cumulative normal distribution
function~\cite{Wasserman04}. Combining the individual property probabilities
can dramatically reduce the probability that the overall alloy will fulfill
the whole specification: for example, if the material has a $50\%$
probability of fulfilling each of the ten specified design criteria, the
overall probability that is fulfills all criteria is $0.5^{10}\approx0.001$,
so $0.1\%$.  Therefore, it is crucial that the probability of the material
meeting the conformance specification is maximized. To achieve this, the
logarithm of the probability $\log(P_{\vec{C}})$ is maximized to ensure
that, in the region where the material is predicted to not satisfy the
specification, the optimizer runs up a constant gradient slope that
persistently favors the least optimized property.

The design tool's use of uncertainty and probability is vital here: the
further the composition is from existing experimental data or the greater
the uncertainty in the experimental data, the larger the uncertainty. The
tool can therefore be allowed to explore the entire range of compositions,
and as soon as it is extrapolating far beyond any available experimental
data points the uncertainty will grow, naturally bounding the range of
compositions from which new alloys may be reliably predicted. The use of
likelihood also allows the tool to explore and select the ideal compromise
between material properties, which is inaccessible to methods that
  do not account for likelihood such as a principal component
  analysis~\cite{Toda13} and robust design~\cite{Backman06}.
Similarly, the design tool may interpolate between experimental
  data, exploring more compositions than would be accessible by an
  experimentally driven search. Using a neural network to interpolate allows
  us to capture deeper correlations than linear regression methods such as a
  principal component analysis~\cite{Toda13}.

As well as predicting material properties, the tool must optimize them.
Previous optimization techniques included running over a pre-determined grid
of compositions, and then sieving them with orthogonal
~\cite{Joo09,Xu09,Kuehmann09,Reed09,Tancret13}, or a Pareto
set~\cite{Bligaard03,Greeley06,Lejaeghere13}. However the expense of these
methods scales exponentially with the number of design variables rendering
them impractical. Another approach is the use of genetic
algorithms~\cite{Johannesson02,Stucke03}. However, this approach is not
mathematically guaranteed to find the optimal
solution~\cite{Ingber92,Mahfoud95}, and it displays poor performance in high
dimensional problems~\cite{Ingber92,Mahfoud95}. Both of these disadvantages
can be overcome by adopting a simulated annealing based approach. With this
approach, a step length can be used that is comparable to the accuracy with
which a material could be manufactured, this is approximately
$0.1\,\rm{wt}.\,\%$ for each element within the entire composition
  excluding the possibility of microsegregation. The temperature is
selected so that a fraction of 0.267 steps are accepted, the optimum to
explore a multidimensional space~\cite{Gelman96}. If too few steps are being
accepted the tool reduces the temperature parameter, if too many then
temperature increases.  This allows the tool to rapidly explore the design
variable space whilst not getting stuck in local minima. Using this
approach, searches of over $\sim10^8$ sets of design variables are
accomplished in $\sim1$ hour to identify an optimal material.

\section{Results and discussion}

Nickel-base superalloys display remarkable high temperature mechanical
properties and environmental resistance. They are used in the gas turbine
engines of the aerospace, marine, and power generation industries, meaning
that there is significant commercial and environmental motivation to develop
improved alloys to enable more efficient engines to be designed and
emissions reduced~\cite{Halse04}. The demanding operating conditions at the
heart of an engine mean that a superalloy must simultaneously satisfy at
least the eleven target properties shown in \tabref{tbl:DataSources}. As
nickel-base superalloys have been intensely studied since the
1930's~\cite{Sims1984,reed2009,thermocalcgeneral02,connor2009,specialmetals2013,haynesinternational2013,
  tomasello1996,radavich2004,sims1987,ewing1976,sato1993,mannan2000,mitchell2004,meng1984,tien1989,
  huron2004,sharma1983,cowen2008,pike2008,mannan2004,tillack1991,rizzo1991,moll1971,brinegar1984,
  bouse1989,braun1989,chang1989,jackman1991,schirra1991,guo1991,mannan2003,rizzo1968,radavich1984,
  xie1996,loewenkamp1988,quested1988,seib2000,shaw1980,sjoeberg2004,nganbe2009,kaufman1984,eng1980,
  brinegar1984_2,specialmetals1971,locq2000,barker1972,hunt2001,wanner1992,tien1990,gu2008,couturier2004,
  sczerzenie1984,jain2000,green1996,furrer2000,ferrari1976,raisson1976,richards1968,maurer1980,reed2010}
and there are over 120 commercially available alloys for different
specialist applications, the optimizing program showed that there is no
single composition that offers properties that exceeds all of the alloys
available. Instead the materials design formalism offers the capability of
finding the ideal compromise between different properties. To demonstrate
this capability, an alloy has been sought that improves the real-life
limiting factors~\cite{Encinas08} of yield stress at 800{\degreeC} and
oxidation resistance, without greatly sacrificing any other property.
The majority of properties are set by the engineering requirements
of the application and to exceed the properties of RR1000 and
Udimet720. For example, elongation is set at the lower bound of the
elongation of Udimet720 of $11.6\%$, as the alloy only needs a sufficient
ductility for damage tolerance in a turbine disc alloy, so this target is
set to allow additional freedom to achieve other property targets. The
$\gamma'$ phase content target is set to control the solvus temperature
below 1150{\degreeC}, alloys which exceed this temperature such as Alloy
10 are known to be difficult to process and suffer from quench
cracking~\cite{Gayda02}.  The alloy has then been benchmarked against the
contemporary alloys Udimet720 (Rolls-Royce AE3007\texttrademark{} engine),
LSHR (NASA proposed disc alloy), Rene104 (General Electric GEnx engine), and
RR1000 (Rolls-Royce Trent 1000 engine).

Models for the properties shown in \tabref{tbl:DataSources} were
constructed, using data from the references highlighted. The yield stress
model focused on arc melted alloys that are readily experimentally
accessible -- in commercial disc applications the alloy would be powder
processed, which typically leads to a $\sim20\%$ higher yield stress through
better grain size and homogeneity control~\cite{hunt2001}. Phase
stability was defined by the volume fraction of the desirable $\gamma$ and
$\gamma'$ phases present after exposure at {$750$\degreeC} for
1000\,hours, this is equivalent to compositions with a value of
$M_{\rm{d}}<0.98$eV that are likely to remain mostly free of undesirable
phases at exposures of {$750$\degreeC} for 1000 hours. Fatigue performance
was taken to be the number of cycles the sample could withstand of a
500\,MPa peak stress oscillating at 60\,Hz. Rupture was the maximum stress
that the alloy can withstand at {$700$\degreeC} for 300 hours. Cr activity
was $\exp[(\mu-\mu_{\rm{Cr}})/RT]$ where $\mu_{\rm{Cr}}$ is the chemical
potential of Cr in the alloy and which ultimately forms a Cr$_2$O$_3$ layer
that protects against further oxidation and corrosion~\cite{reed2010}.

\begin{figure}
 \centering
 \includegraphics[width=1\linewidth]{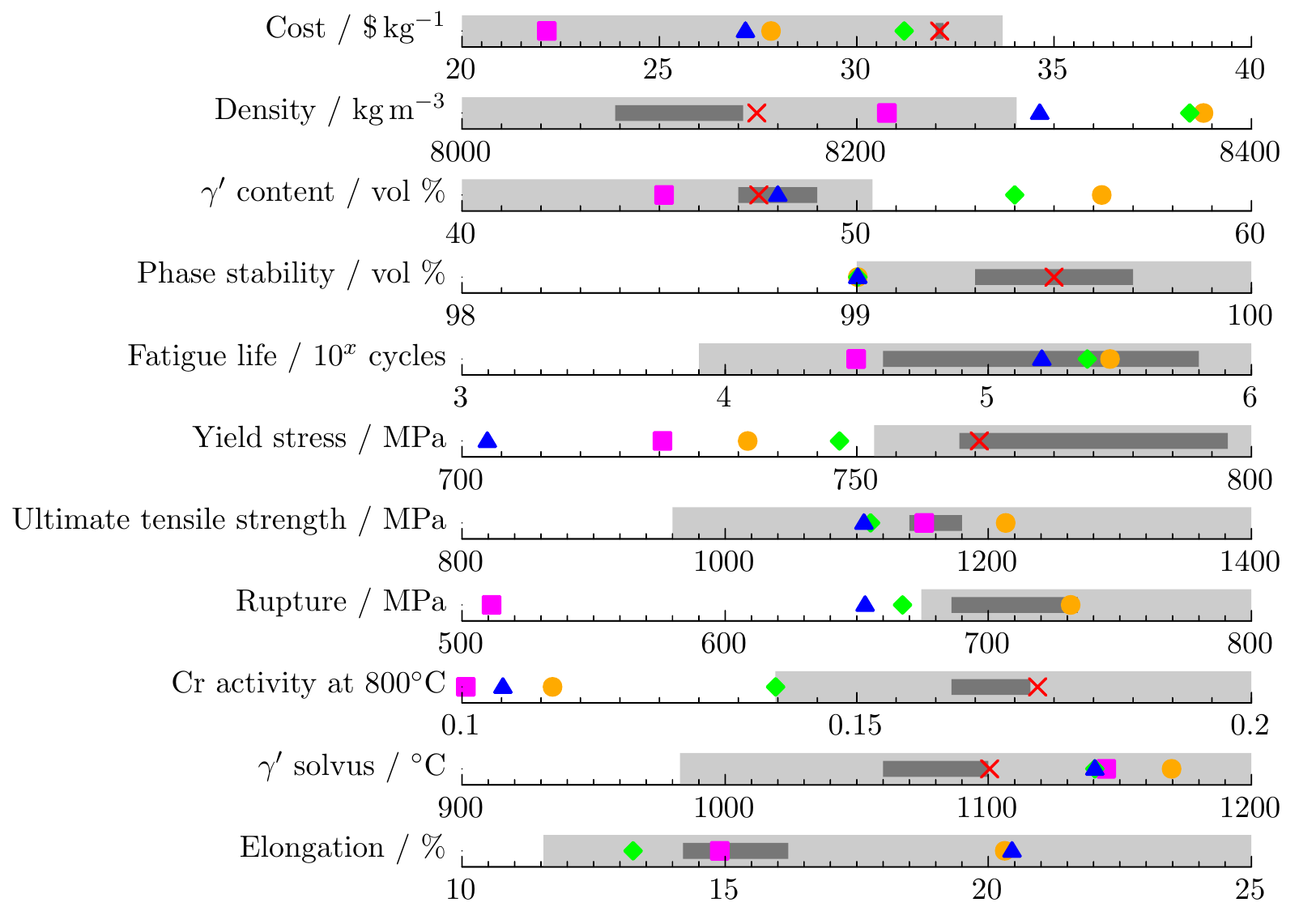}
 \caption{Summary of properties for the Ni superalloy. For each listed property the
   gray box refers to the acceptable target properties, the dark gray is the 
   three-sigma uncertainty on the theoretical
   prediction.
   The points refer to experimentally measured values with
   {\color{red}$\times$}~V210A where measured,
   {\color{magenta}$\blacksquare$}~refers
   to Udimet720, {\color{orange}$\bullet$}~LSHR,
   {\color{green}$\blacklozenge$}~Rene104, and
   {\color{blue}$\blacktriangle$}~RR1000.
 }
 \label{fig:ResultsNi}
\end{figure}

\begin{table}
 \small
 \centering
 \begin{tabular}{ll|ll}
  \multicolumn{4}{l}{Optimal\,composition\,(wt.\,\%)}\\
  \hline
  Mn&\pme{ 0.2}{0.2}&Cr&\pme{15.8 }{0.7}\\
  Co&\pme{20.0}{0.9}&Ti&\pme{3.0  }{0.2}\\
  Mo&\pme{ 0.5}{0.3}&Fe&\pme{3.9  }{0.4}\\
  W &\pme{ 0.5}{0.3}&Si&\pme{0.2  }{0.2}\\
  Ta&\pme{ 4.9}{0.3}&Zr&\pme{0.18 }{0.03}\\
  Nb&\pme{ 1.1}{0.2}&B &\pme{0.06 }{0.01}\\
  Al&\pme{ 2.4}{0.2}&C &\pme{0.02}{0.01}\\
  Ni&Balance        &&\\
  $T$&900\degreeC&$t$&30 hr\\
 \end{tabular}
 \caption{The proposed nickel-base superalloy composition (wt.\,\%) and heat treatment routine with the design
   tolerance of all design variables that are predicted to fulfill the target specification.}
  \label{fig:NiComposition}
\end{table}

The target properties for the Ni superalloy are shown in
\tabref{tbl:DataSources}. We focus on searching for an alloy that has a
better high-temperature yield strength, to allow it to be used in the next
generation of engines with higher operating temperature and also better
oxidation resistance than previous alloys to increase engine service
intervals.  Therefore, an alloy is sought that has a target yield strength
that exceeds $752$\,MPa, greater than the best in the sample set,
Rene104. The alloy must also have higher levels of chromium activity than
previous alloys to promote improved oxidation resistance.  The targets were
set so that the alloy should then be competitive on all other properties.
These models allowed a composition to be proposed along with an appropriate
heat treatment schedule that has a $20\%$ probability of fulfilling the
target specification.  The theoretical predictions, shown in
\figref{fig:ResultsNi}, all fall within the required target specification.
This new alloy is denoted V210A, the composition of which is given in
\tabref{fig:NiComposition}.

\subsubsection{Analysis of proposed alloy}

The aim of the alloy design tool is to select the alloy with the highest
probability of exceeding all the targets. Examining the probability of an
alloy satisfying the design targets exposes the compromise that has been
made between the physical properties stress rupture, phase stability,
thermal expansivity, yield stress, and ultimate tensile strength.

\begin{figure}
 \centering
 \begin{tabular}{ll}
  (a)\\\includegraphics[width=0.9\linewidth]{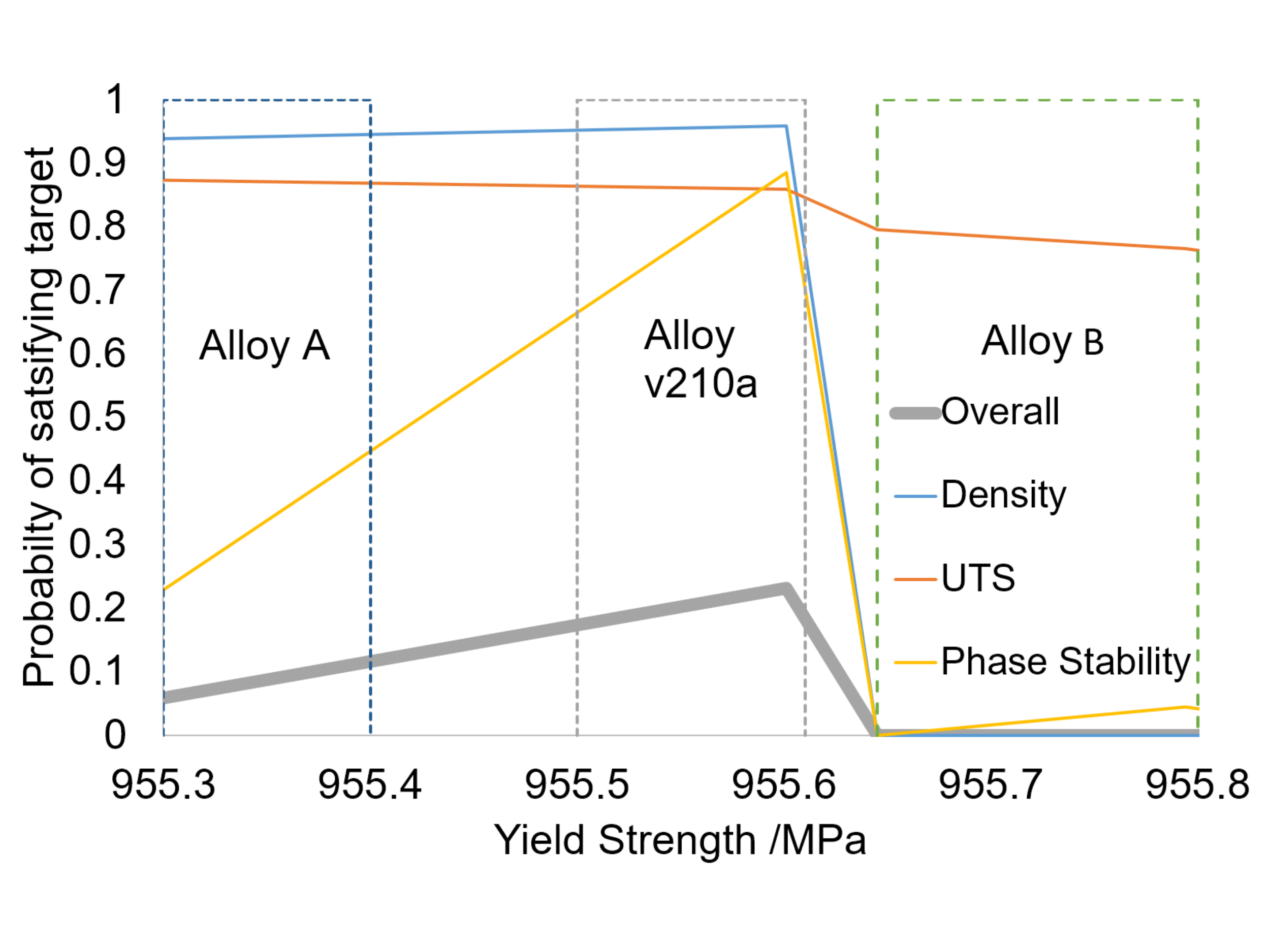}\\
  (b)\\\includegraphics[width=0.9\linewidth]{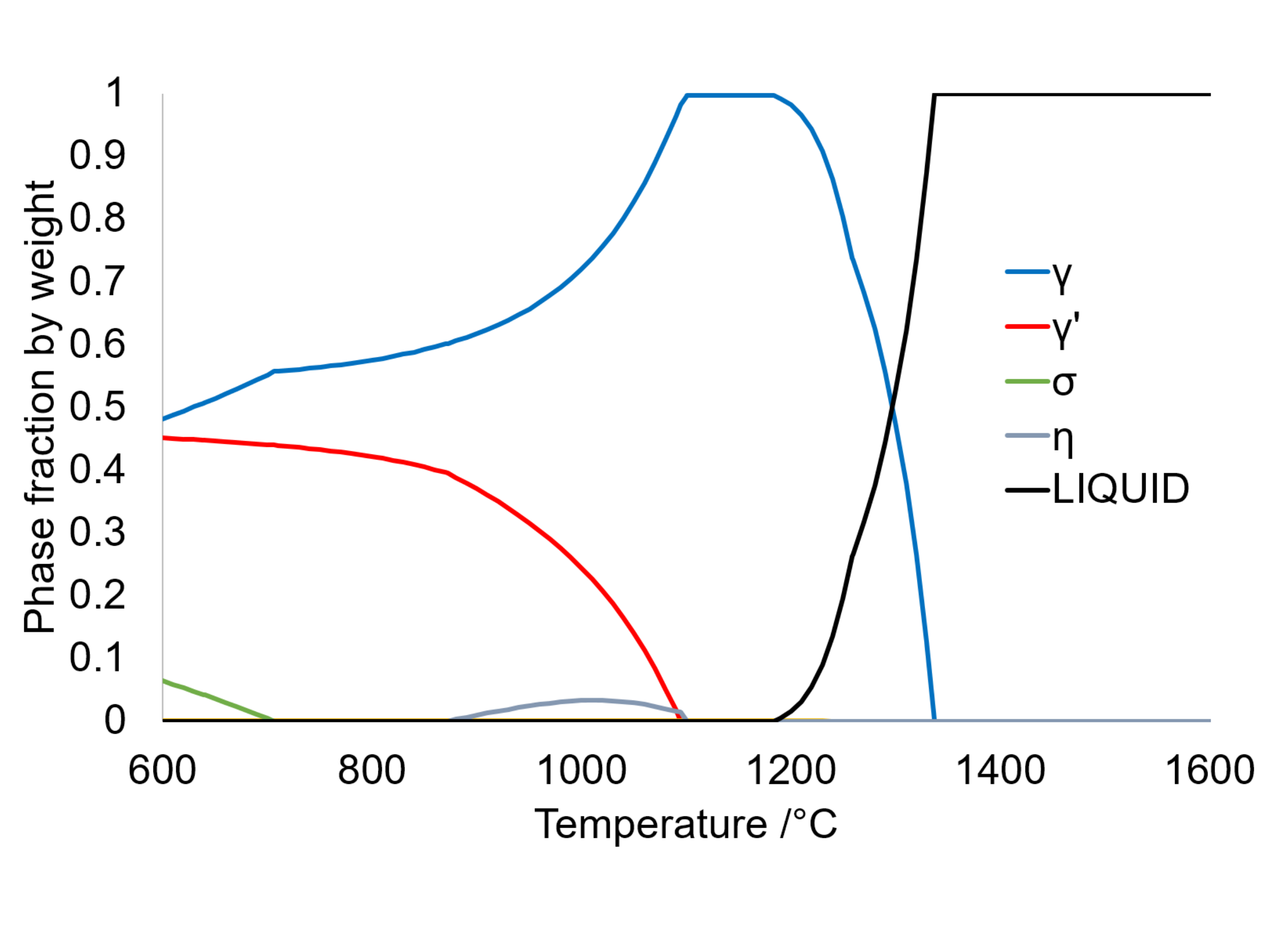}
 \end{tabular}
 \caption{(Color online)
   (a) The variation of the probability of satisfying the overall (gray),
   density (blue), ultimate tensile strength (red), and
   phase stability target (yellow) as a function of yield strength
   of the alloy. As well as V210A, two
   additional alloys, A and B, are described in the text.
   (b) The phase fraction of the $\gamma$ (blue), $\gamma'$ (red), $\sigma$ (green),
   $\eta$ (gray), and liquid (black) phases for the alloy V210A.
 }
 \label{fig:YSCompromise}
\end{figure}

An example of the compromises that must be made between properties is
presented in \figref{fig:YSCompromise}(a) in which the probability of three
compositionally similar alloys (Alloy A, Alloy B, and Alloy V210A) exceeding
various property targets are shown as a function of yield stress. All three
alloys are predicted to have a high yield stress, in excess of the target
specification. However, the probability that they meet the other target
properties varies considerably. Alloy A is predicted to have a high
probability of exceeding the density and ultimate tensile strength targets
but falls short in respect to phase stability. Alloy B has a high
probability of exceeding the UTS target but a poor prediction in terms of
phase stability and density. Alloy V210A has a high probability of exceeding
all three property targets plotted. Thus, when the properties are combined
to give an overall likelihood of meeting all the targets, V210A has a
probability which exceeds both that of Alloy A and Alloy B. It should be
noted that, because the plot shows how the probability varies with a single
property, yield stress, rather than composition, there is a sharp peak in
property space.

Physical properties of Ni alloys are controlled not only by composition but
also with the operating temperature. Therefore, to understand the variation
of physical properties, in \figref{fig:YSCompromise}(b) we examine the
predicted mass fraction of phases as a function of temperature for the alloy
V210A. The $\gamma'$ solvus temperature is predicted to be
$1100${\degreeC}. Between the temperatures of $870$-$1100${\degreeC}, a
small percentage of $\eta$ phase (Ni$_3$Ti) is predicted to form; if the
alloy is processed above the solvus temperature and operated at
$<850${\degreeC} then the volume fraction observed in the alloy should be
close to zero. Out of all the deleterious phases that can be formed
within V210A, the $\sigma$ phase is predicted to form at the highest
temperature, yet it is $<700${\degreeC}; below these temperatures the
kinetics of diffusion within the alloy will be slow enough to prevent
formation of this phase. The $\sigma$ phase forms a "basket weave"
morphology, which has a severe impact on mechanical properties that
affects many commercially available alloys. The rapidly varying phase
stability of the alloy, as indicated in \figref{fig:YSCompromise}(a), means
that the alloy properties are expected to vary rapidly across property
space.

Having seen how the yield stress, phase stability, ultimate tensile
strength, and density can act in counter directions it is appropriate to
examine how the probabilities of all the properties vary in
tandem. Therefore, pairs of properties have been selected and examined to
elucidate the compromise that must be made between them. For each pair, two
graphs are presented, the first graph shows a relief plot of the probability
of exceeding the two targets; the second shows the probability of exceeding
all the other targets that are taken into consideration.

\begin{figure}[t]
 \centering
 \begin{tabular}{ll}
   (a)\\\includegraphics[width=0.9\linewidth]{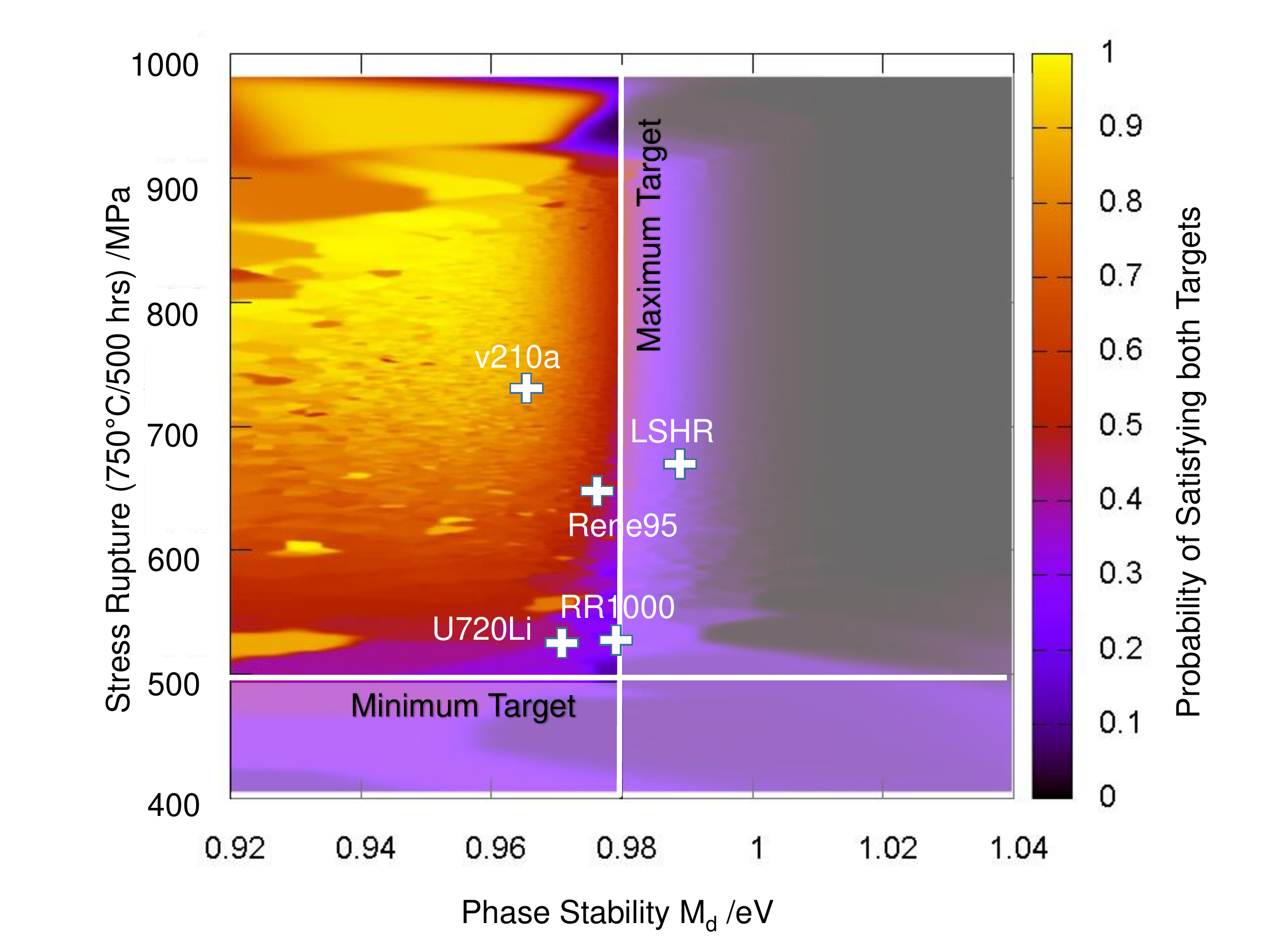}\\
  (b)\\\includegraphics[width=0.9\linewidth]{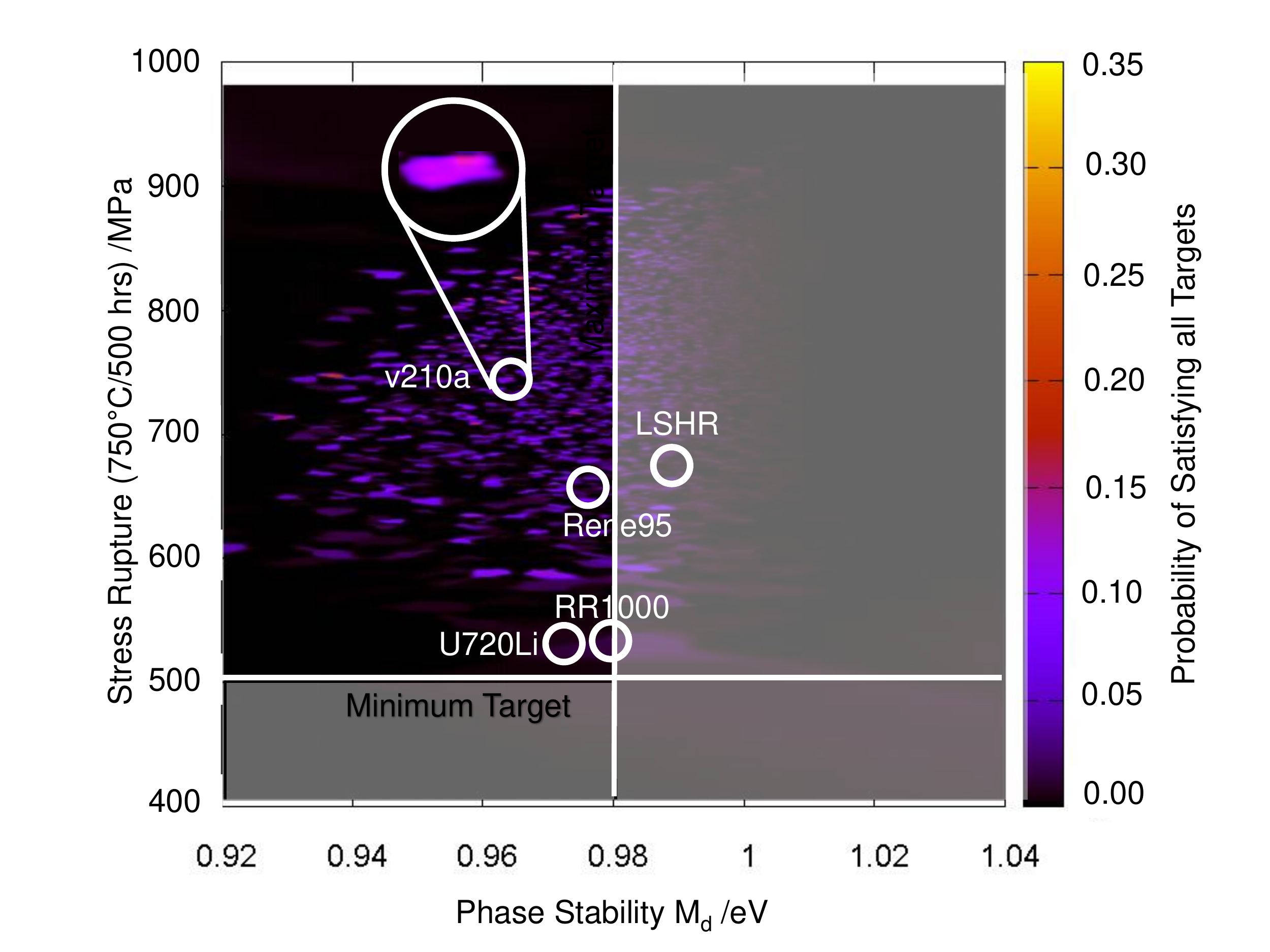}
 \end{tabular}
 \caption{(Color online)
   (a) The probability of simultaneously satisfying two targets, a maximum
   phase stability target of $0.98$\,eV and a minimum stress rupture target
   of $500$\,MPa at $750${\degreeC} for 100 hours.
   (b) A relief plot of simultaneously satisfying all of the targets for
   the material, plotted with respect to phase stability and
   minimum stress rupture targets.
 }
 \label{fig:StressRupturePhaseStability}
\end{figure}

{\it Stress rupture and phase stability}:
\figref{fig:StressRupturePhaseStability}(a) shows the predicted probability
of simultaneously satisfying the minimum targets for stress rupture and
phase stability. A minimum target for a stress rupture resistance of
$500$\,MPa at $750${\degreeC} for $100$\,hours and a maximum phase
stability, $M_{\rm{d}}=0.98$\,eV was set. This was set by comparison of
the $M_{\rm{d}}$ values and $\sigma$ phase formation with existing
superalloys.

The regions of high probabilities (bright colors) signify good predicted
properties coupled with a low uncertainty due to the higher density of
historical data-points around these regions. The lowest probabilities are in
the regions where neither target is predicted to be satisfied or where there
is a low data density leading to high uncertainty. The rapidly varying
pockets of high and low probability are driven by the different properties
coming in and out of favor and the phases present vary rapidly across the
composition space, as we saw in \figref{fig:YSCompromise}. Since the targets
were defined using a normal distribution, at the point $500$\,MPa, and
$0.98$\,eV the probability of simultaneously satisfy both targets is
$0.25$. In general, the highest probabilities are around a stress of
$850$\,MPa, and phase stability of $0.95$\,eV. Creating an alloy with a
higher stress rupture resistance and lower phase stability than this becomes
increasingly unlikely.

In general, an increase in stress rupture resistance requires additional
alloying elements, which results in a decrease in the phase
stability~\cite{Morinaga84}. Our probability based approach enables the best
compromise between these two competing factors to be found. In addition, it
helps establish which elements offer the most effective way of increasing
the probability of exceeding the stress rupture target whilst resulting in
only a modest decrease in phase stability.

The proposed alloy, V210A, is contained within the large region of high
probability well above the minimum stress target, whilst RR1000 and U720Li
are just above the target. Rene95 and LSHR have superior stress rupture
resistance, but neither Rene95 nor LSHR meet the phase stability
target. This graph only shows the probabilities for two properties, most of
the high probability region above V210A become infeasible when the other
properties are taken into account.

\begin{figure}[t]
 \centering
 \begin{tabular}{ll}
  (a)\\\includegraphics[width=0.9\linewidth]{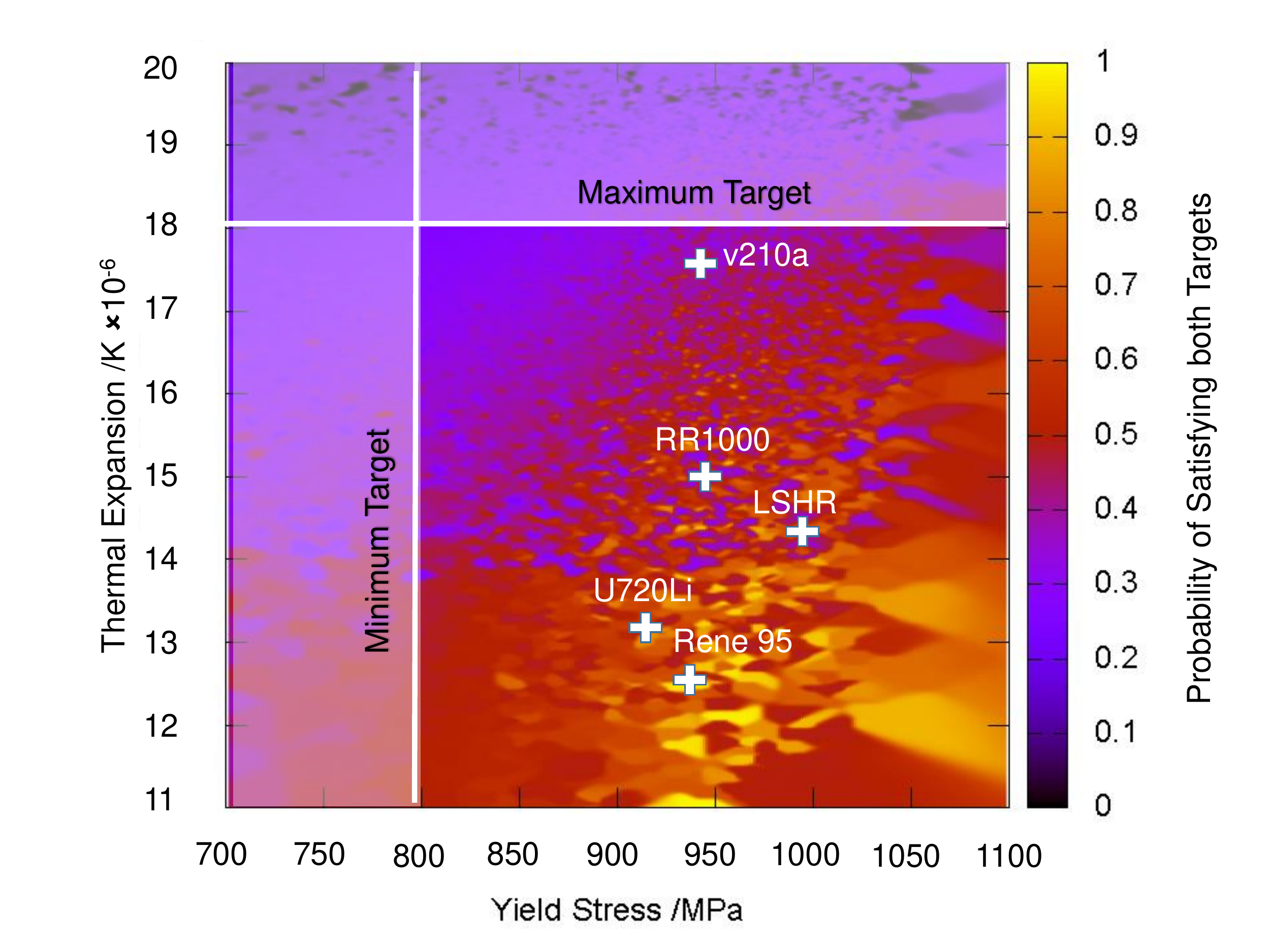}\\
  (b)\\\includegraphics[width=0.9\linewidth]{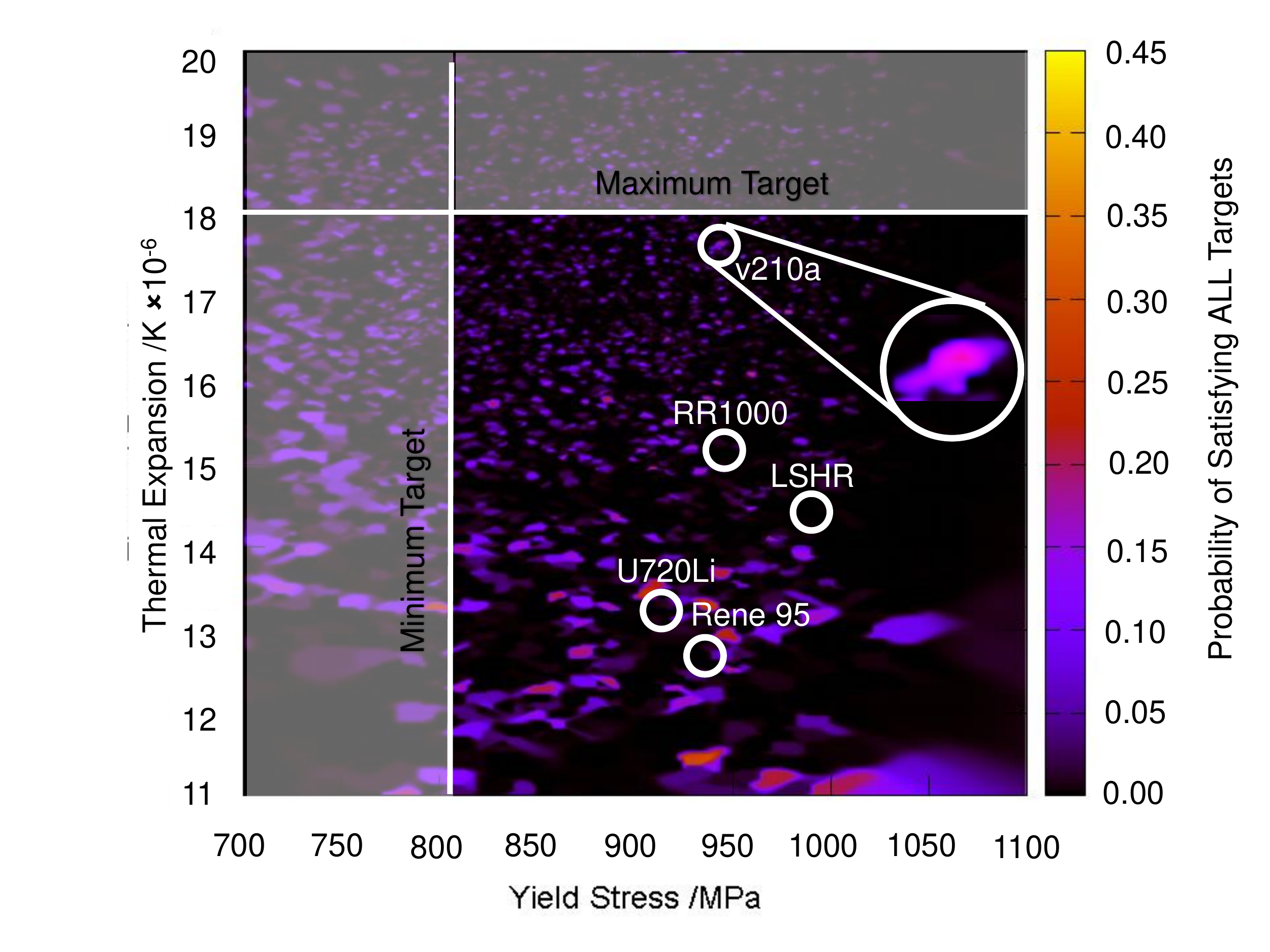}
 \end{tabular}
 \caption{(Color online)
   (a) A relief plot of simultaneously satisfying two targets, a maximum
   thermal expansion target of $17\times10^{-6}/$K at $750${\degreeC} and a
   minimum yield stress of $800$\,MPa at $750$\degreeC.
   (b) A relief plot of simultaneously satisfying all targets for the
   material, plotted with respect to thermal expansion and yield stress targets.
 }
 \label{fig:ThermalExpansivityYS}
\end{figure}

\begin{figure}[t]
 \centering
 \begin{tabular}{ll}
  (a)\\\includegraphics[width=0.9\linewidth]{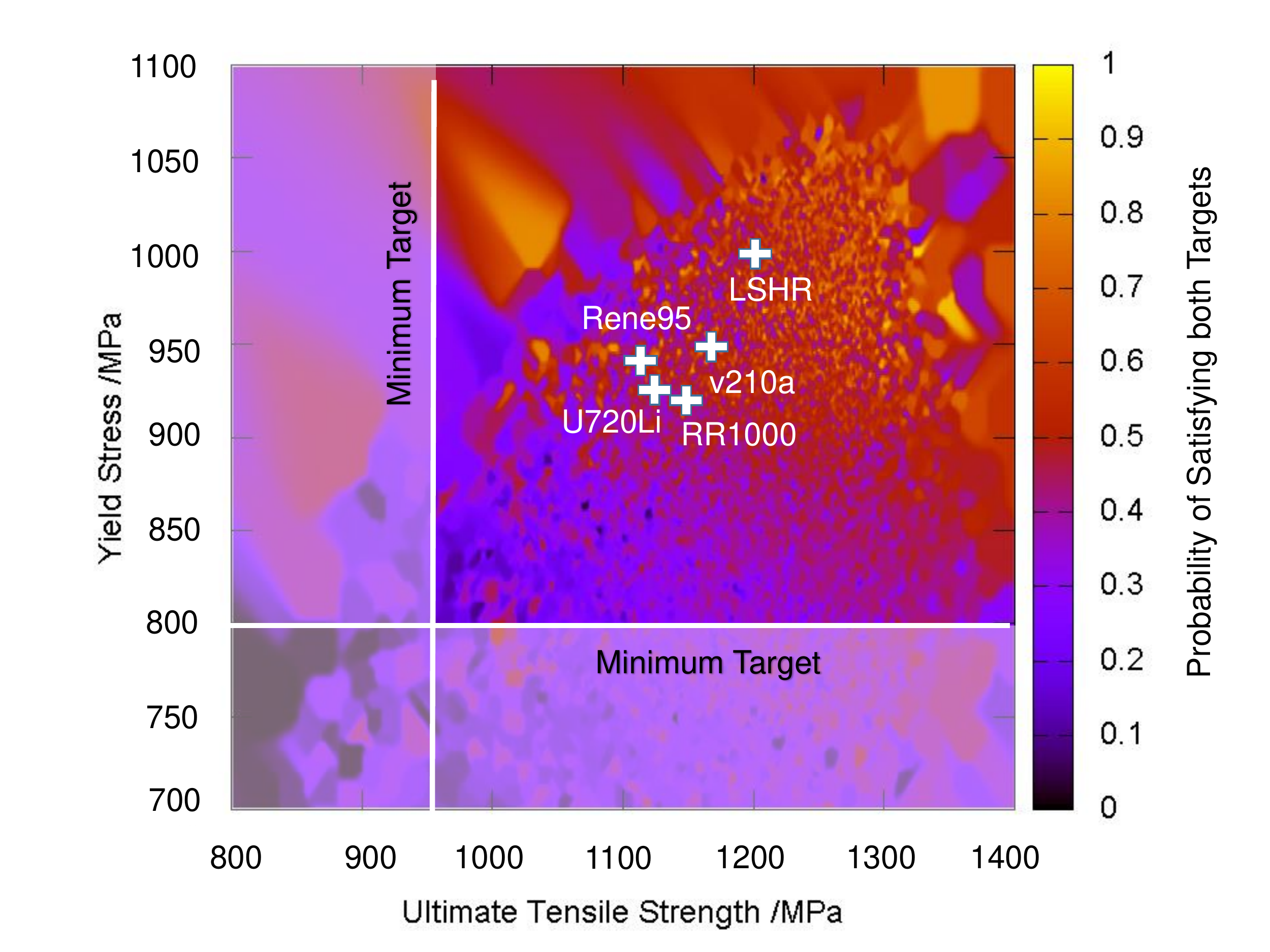}\\
  (b)\\\includegraphics[width=0.9\linewidth]{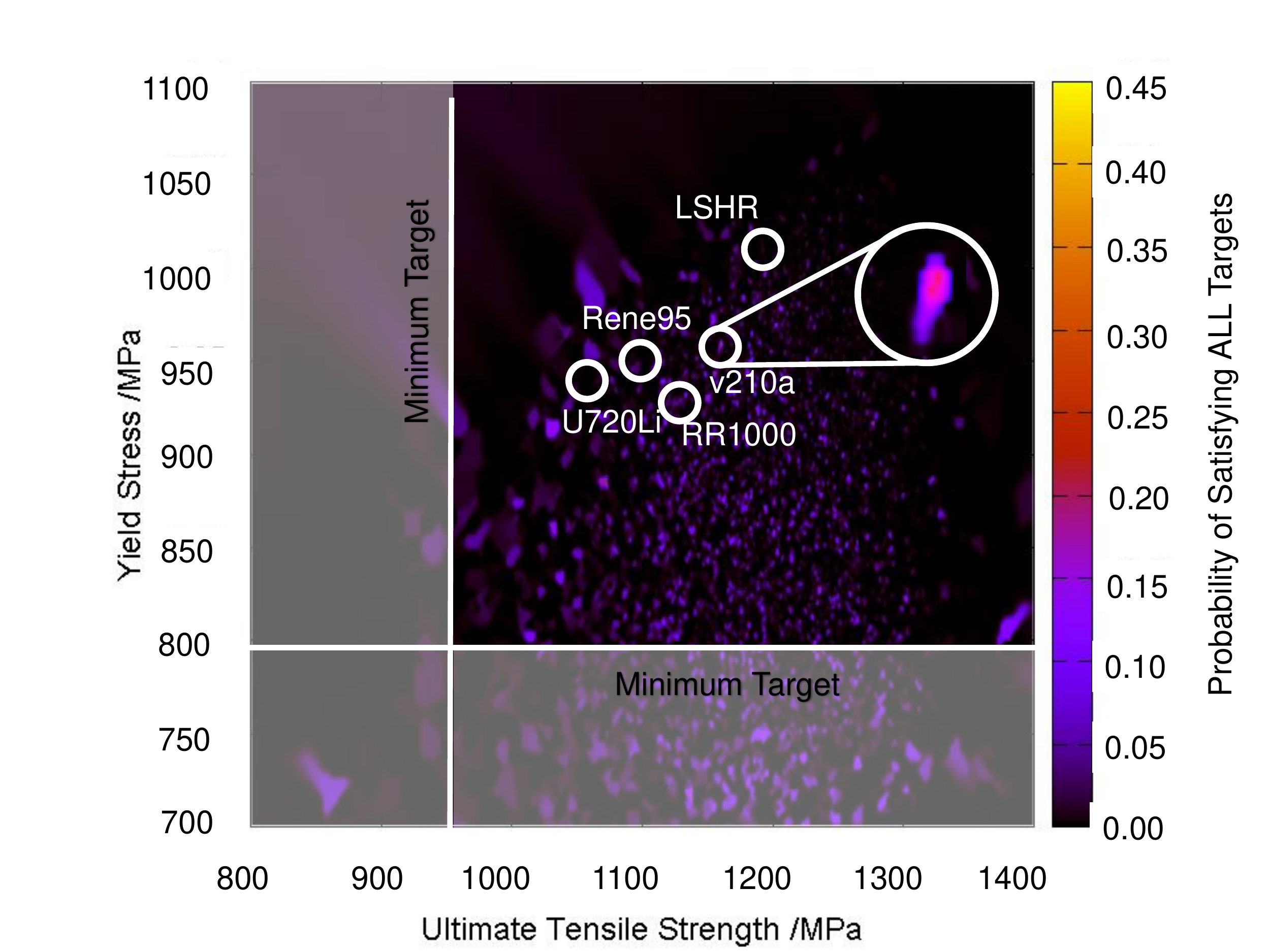}
 \end{tabular}
 \caption{(Color online)
   (a) A relief plot of simultaneously satisfying two targets,
   a minimum yield stress target of $850$\,MPa at $750${\degreeC}
   and a minimum ultimate stress target of $950$\,MPa at $750$\degreeC.
   (b) A relief plot of simultaneously satisfying all targets for the
    material, plotted with respect to yield stress and ultimate tensile stress targets.
 }
 \label{fig:YSUTS}
\end{figure}

\figref{fig:StressRupturePhaseStability}(b) shows how the probability
distribution with respect to stress rupture and phase stability change when
all the other property targets are taken into account. This drastically
changes the distribution of probabilities shown in
\figref{fig:StressRupturePhaseStability}(a). Now there are sparse regions of
higher probability of exceeding all targets from that obtained when only two
properties are considered. V210A is contained within a region of high
probability, which we confirm is not noise; firstly, because all of the
relief plots, \figref{fig:StressRupturePhaseStability},
\figref{fig:ThermalExpansivityYS}, and \figref{fig:YSUTS} all show V210A in
a region of high probability, secondly because the consideration of
probability means that noise has large uncertainty and so would lower the
probability, and thirdly by our later experimental verification.  The figure
correctly shows that the alloys LSHR, Rene95, RR1000, U720Li do not fulfill
the targets specified. Having seen the impact of the inclusion of the other
properties, we next look at the variation of design probability with thermal
expansivity and yield stress.

{\it Thermal expansivity and yield stress}:
\figref{fig:ThermalExpansivityYS}(a) shows a relief probability plot of
thermal expansivity coefficient as a function of yield stress for developing
the alloy V210A. Minimum targets of a yield stress of $800$\,MPa at
$750${\degreeC} and a maximum thermal expansion coefficient of
$17\times10^{-6}/$K at $750${\degreeC} were set. The region with the highest
probability of exceeding these two targets is in general at the lowest
thermal expansion coefficient and highest yield stress. The thermal
expansion coefficient can be improved by adding refractory elements such as
Mo, Nb, W and Ta, which will simultaneously improve the yield
stress. However, adding these elements will increase the density and reduce
phase stability, which makes the majority of the family of alloys
infeasible, as shown in the previous graphs.

\figref{fig:ThermalExpansivityYS}(b) shows how the probability
distribution with respect to thermal expansion and yield stress changes when
all the other property targets are taken into account.  There is not a
significant increase on the probability of success with decreasing thermal
expansion. Therefore, it may be better to choose an alloy with a relatively
high thermal expansion coefficient, as is the case for V210A -- which has the
highest probability of exceeding all the targets out of the alloys
shown. The other alloys shown have similar yield stresses and lower thermal
expansion coefficient, but as highlighted previously do not have all the other
necessary requirements for the application in question.

{\it Yield stress and ultimate tensile strength}: \figref{fig:YSUTS}(a)
shows a relief plot of simultaneously satisfying two targets, a minimum
yield stress target of $800$\,MPa at $750${\degreeC} and a minimum ultimate
stress target of $950$\,MPa at $750${\degreeC}. \figref{fig:YSUTS}(b) shows
how the probability changes if all targets are considered. Now there are
many small regions with higher probability of success surrounded by larger
regions with a low probability of success. The distribution of probability
regions is similar to that seen in the yield stress vs. thermal expansion
coefficient.

As can been observed from all of these plots, V210A, sits in a region of
high probability which the other alloys do not, and thus is the alloy most
likely to meet the targets specified. Obviously, a different set of targets
would result in a different distribution of high probability alloys. This
highlights the potential advantages of matching the alloy for the specific
application.

\subsubsection{Experimental verification}

Experimental testing was performed to verify the beneficial properties of
the proposed alloy, V210A. Starting from pelletized elements with purity
greater than $99.9\%$, the mixture was arc-melted under argon to produce a
$50\,$g ingot through five successive inversion and re-melt cycles. The
alloy was then homogenized for $72\,$hours at {$1200$\degreeC}, followed by
the heat treatment in \tabref{fig:ResultsNi} of holding at {$900$\degreeC}
for $30\,$hours. The alloy's yield stress was determined through compression
testing on 4\,mm diameter, 6\,mm long cylindrical samples. Following a
15\,minute dwell at the testing temperature a strain rate of 0.001\,s$^{-1}$
was applied to measure the 0.2\% proof strength. To study the oxidation
resistance, a $20\times10\times0.5$\,mm sample, whose surface was ground
with 3\,\textmu{}m grit paper was subjected to thermo-gravimetric analysis at
{$800$\degreeC} in air.

\begin{figure}[t]
 \centering
 \includegraphics[width=0.65\linewidth]{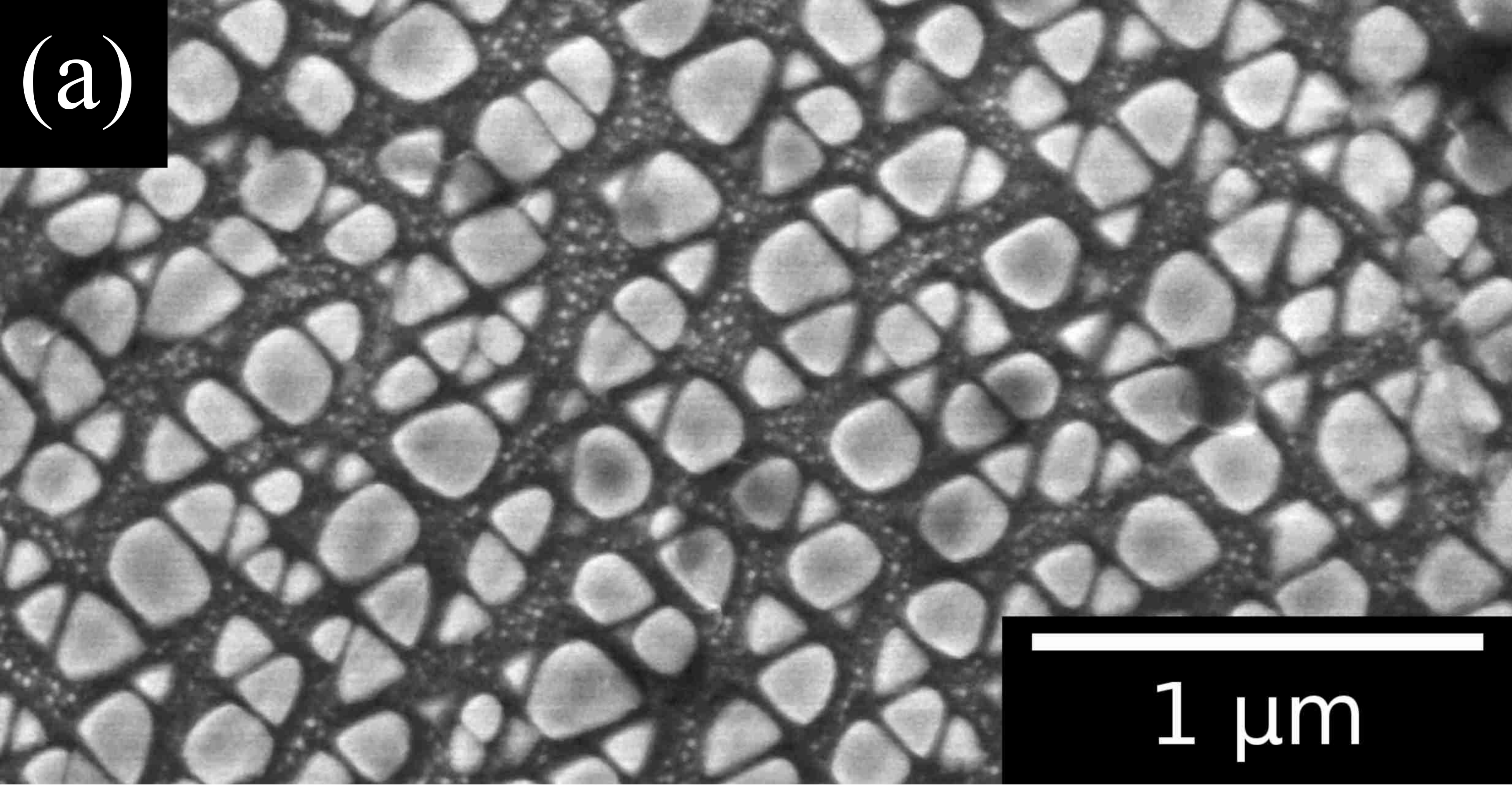}\\\vspace{6pt}
 \includegraphics[width=0.65\linewidth]{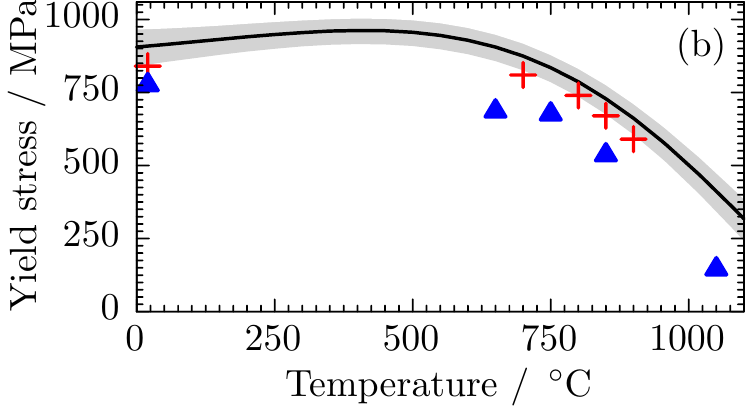}\\\vspace{6pt}
 \includegraphics[width=0.65\linewidth]{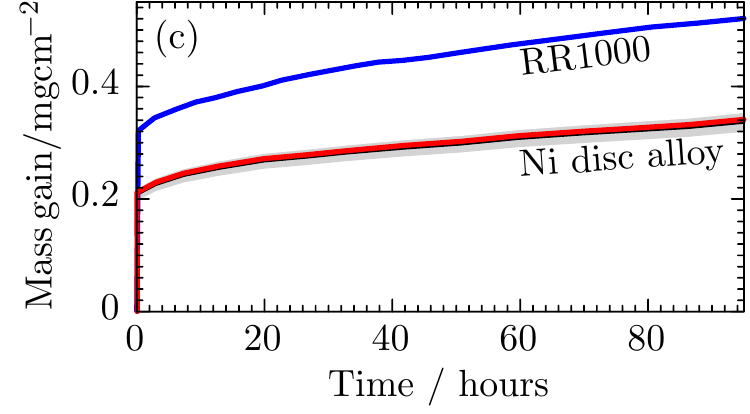}
 \caption{(Color online)
   (a) Secondary electron micrograph image.
   (b) V210A yield stress as a
   function of temperature with black the theoretical prediction for the
   proposed alloy, along with the uncertainty in gray. The points
   {\color{red}$\boldsymbol{+}$}~show experimental results for the optimal
   alloy and {\color{blue}$\blacktriangle$}~RR1000.
   (c) Oxidation resistance of V210A and RR1000 with
   temperature. The theoretical predictions are shown in black with
   uncertainty in gray.
 }
 \label{fig:ExperimentNi}
\end{figure}

The predictions of the phase behavior were verified through secondary
electron microscopy, as shown in \figref{fig:ExperimentNi}(a). The measured
volume fraction of $\gamma'$ precipitates was determined to be $51\%$,
confirming our predictions, and less than 1\% of other deleterious phases
were identified.  The $\gamma'$ solvus temperature was determined using
metallographic techniques. A small series of ingots were prepared and
annealed at $\pm10${\degreeC} of the predicted solvus temperature.  The
samples were quenched and examined under electron microscopy.  Examination
of the morphologies of $\gamma'$ present confirmed that the solvus
temperature lay within $10${\degreeC} of the predicted result,
$1100${\degreeC}. The experimentally measured density, determined by the
Archimedes method, also matched the theoretical
prediction. \figref{fig:ExperimentNi}(b) shows that the theoretical and
experimental yield stress for an arc melted alloy agree within expected
uncertainty, and exceed experimentally measured values of the commercial
alloy RR1000. The proposed alloy performs well at high temperature due to
the retention of $\gamma'$ precipitate strengthening, as well as solid
solution strengthening. For a powder processed alloy, reduced grain size
means that the yield stress would be even greater.
\figref{fig:ExperimentNi}(c) shows that the oxidation resistance agrees with
the theoretical prediction and is superior to that of RR1000.

Having examined the experimental results we can summarize and compare all of
the seven properties measured (cost, density, $\gamma'$ content, phase
stability, yield stress, Cr activity, and $\gamma'$ solvus) to the model
predictions and the values reported for other commercially available alloys
in \figref{fig:ResultsNi}. The properties of V210A are consistent with
theoretical predictions, within uncertainty. In particular, V210A exceeds
the required targets of yield stress and oxidation resistance, that none of
the exemplar commercial alloys Udimet720, LSHR, Rene104, and RR1000
simultaneously fulfill. Moreover, the new alloy design tool allows us to see
how previous alloys do not have the appropriate compromise between
properties, for example Udimet720 is a low cost, lightweight alloy with low
$\gamma'$ content, but it has comparatively poorer stress rupture, yield
stress, and oxidation resistance. On the other hand, LSHR fulfills the
stress rupture target, but the density and $\gamma'$ content are arguably
too high, along with insufficient oxidation resistance. The ability of the
neural network tool to optimize all material properties simultaneously means
that the proposed nickel-base superalloy is an ideal candidate for its target
application.

\section{Conclusions}

A new computational alloy design tool was developed that incorporates
uncertainty to allow alloys to be designed with the greatest probability of
meeting a design specification containing many different material
properties. The design tool was used to propose a new nickel-base superalloy
alloy most likely to simultaneously fulfill eleven different physical
criteria. The tool predicted that the new nickel-base polycrystalline alloy
offered an ideal compromise between its properties for disc applications and
seven of these properties were experimentally verified, demonstrating that
it has better yield stress and oxidation resistance than commercially
available alternatives. The tool has also been used to design a nickel-base
alloy for a combustor liner~\cite{Conduit2013ii}, and two Mo-based alloys
for forging tools~\cite{Conduit2014v,Conduit2014vi}. The capability to
rapidly discover materials computationally using this approach should
empower engineers to rapidly optimize bespoke materials for a given
application, bringing materials into the heart of the design process.

\section*{Acknowledgments}
The authors thank Mark Hardy for useful
discussions and acknowledge the financial support of Rolls-Royce
plc, EPSRC under EP/H022309/1 and EP/H500375/1, and Gonville \& Caius
College. There is Open Access to this paper
at \texttt{https://www.openaccess.cam.ac.uk}.

\bibliographystyle{unsrtnat}
\bibliography{NeuralData,StdPapers}

\begin{thebibliography}{90}
\providecommand{\natexlab}[1]{#1}
\providecommand{\url}[1]{\texttt{#1}}
\expandafter\ifx\csname urlstyle\endcsname\relax
  \providecommand{\doi}[1]{doi: #1}\else
  \providecommand{\doi}{doi: \begingroup \urlstyle{rm}\Url}\fi

\bibitem[Curtarolo et~al.(2013)Curtarolo, Hart, Nardelli, Mingo, Sanvito, and
  Levy]{Curtarolo13}
S.~Curtarolo, G.L.W. Hart, M.B. Nardelli, N.~Mingo, S.~Sanvito, and O.~Levy.
\newblock The high-throughput highway to computational materials design.
\newblock \emph{Nature Materials}, 12:\penalty0 191, 2013.

\bibitem[Kuehmann and Olson(2009)]{Kuehmann09}
C.~Kuehmann and G.B. Olson.
\newblock Computational materials design and engineering.
\newblock \emph{Materials Science and Engineering}, 25:\penalty0 472, 2009.

\bibitem[Bligaard et~al.(2003)Bligaard, J\'ohannesson, Ruban, Skriver,
  Jacobsen, and N{\o}rskov]{Bligaard03}
T.~Bligaard, G.H. J\'ohannesson, A.V. Ruban, H.L. Skriver, K.W. Jacobsen, and
  J.K. N{\o}rskov.
\newblock Pareto-optimal alloys.
\newblock \emph{Appl. Phys. Lett.}, 83:\penalty0 4527, 2003.

\bibitem[Greeley et~al.(2006)Greeley, Jaramillo, Bonde, Chorkendorff, and
  N{\o}rskov]{Greeley06}
J.~Greeley, T.F. Jaramillo, J.~Bonde, I.~Chorkendorff, and J.K. N{\o}rskov.
\newblock Computational high-throughput screening of electrocatalytic materials
  for hydrogen evolution.
\newblock \emph{Nature Materials}, 5:\penalty0 909, 2006.

\bibitem[Lejaeghere et~al.(2013)Lejaeghere, Cottenier, and
  Van~Speybroeck]{Lejaeghere13}
K.~Lejaeghere, S.~Cottenier, and V.~Van~Speybroeck.
\newblock Ranking the stars: A refined pareto approach to computational
  materials design.
\newblock \emph{Phys. Rev. Lett.}, 111:\penalty0 075501, 2013.

\bibitem[Toda-Caraballo et~al.(2013)Toda-Caraballo, Galindo-Nava, and
  Rivera-D\'iaz-del Castillo]{Toda13}
I.~Toda-Caraballo, E.I. Galindo-Nava, and P.E.J. Rivera-D\'iaz-del Castillo.
\newblock Unravelling the materials genome: symmetry relationships in alloy
  properties.
\newblock \emph{Journal of Alloys and Compounds}, 566:\penalty0 217, 2013.

\bibitem[Backman et~al.(2006)Backman, Wei, Whitis, Buczek, Finnigan, and
  Gao]{Backman06}
D.G. Backman, D.Y. Wei, D.D. Whitis, M.B. Buczek, P.M. Finnigan, and D.~Gao.
\newblock Icme at ge: Accelerating the insertion of new materials and
  processes.
\newblock \emph{Journal of Materials}, page 36–41, 2006.

\bibitem[Joo et~al.(2009)Joo, Ryu, and Bhadeshia]{Joo09}
M.~Joo, J.~Ryu, and H.K.D.H Bhadeshia.
\newblock Domains of steels with identical properties.
\newblock \emph{Mater. Manuf. Process.}, 24:\penalty0 53–58, 2009.

\bibitem[Xu et~al.(2009)Xu, Rivera Diaz~del Castillo, and van~der Zwaag]{Xu09}
W.~Xu, P.E.J. Rivera Diaz~del Castillo, and S.~van~der Zwaag.
\newblock A combined optimization of alloy composition and aging temperature in
  designing new uhs precipitaition hardenable stainless steels.
\newblock \emph{Computational Materials Science}, 45:\penalty0 467, 2009.

\bibitem[Reed et~al.(2009{\natexlab{a}})Reed, Tao, and Warnken]{Reed09}
R.C. Reed, T.~Tao, and N.~Warnken.
\newblock Alloys-by-design: application to nickel-based single crystal
  superalloys.
\newblock \emph{Acta Materialia}, 57:\penalty0 5898, 2009{\natexlab{a}}.

\bibitem[Tancret(2013)]{Tancret13}
F.~Tancret.
\newblock Computational thermodynamics, gaussian processes and genetic
  algorithms: combined tools to design new alloys.
\newblock \emph{Modelling Simul. Mater. Sci. Eng.}, 21:\penalty0 045013, 2013.

\bibitem[Andersson et~al.(2002)Andersson, Helander, H\"oglund, Shi, and
  Sundman]{thermocalcgeneral02}
J.O. Andersson, T.~Helander, L.~H\"oglund, P.F. Shi, and B.~Sundman.
\newblock {Thermo-Calc and DICTRA, Computational tools for materials science}.
\newblock \emph{Calphad}, 26:\penalty0 273--312, 2002.

\bibitem[Conduit and Conduit(2014)]{Conduit2014iv}
B.D. Conduit and G.J. Conduit.
\newblock Method and system for designing a material.
\newblock Patent EP14153898, US 2014/177578, 2014.

\bibitem[Conduit et~al.(2014{\natexlab{a}})Conduit, Conduit, Stone, and
  Hardy]{Conduit2014vii}
B.D. Conduit, G.J. Conduit, H.J. Stone, and M.C. Hardy.
\newblock A nickel alloy.
\newblock Patent EP14157622, amendment to US 2013/0052077 A2,
  2014{\natexlab{a}}.

\bibitem[Morinaga et~al.(1984)Morinaga, Yukawa, Adachi, and Ezaki]{Morinaga84}
M.~Morinaga, N.~Yukawa, H.~Adachi, and H.~Ezaki.
\newblock New phacomp and its application to alloy design.
\newblock In \emph{Superalloys 1984}, pages 523--532. The Minerals, Metals \&
  Materials Society, 1984.
\newblock ISBN 9780895204783.

\bibitem[Kaufman and Bernstein(1970)]{Kaufman70}
L.~Kaufman and H.~Bernstein.
\newblock \emph{Computer Calculation of Phase Diagrams}.
\newblock Academic Press, 1970.

\bibitem[Heskes(1997)]{Heskes97}
T.~Heskes.
\newblock Practical confidence and prediction intervals.
\newblock In \emph{Advances in Neural Information Processing Systems 9}, pages
  176--182. MIT press, 1997.
\newblock ISBN 9780262100656.

\bibitem[Papadopoulos et~al.(2001)Papadopoulos, Edwards, and
  Murray]{Papadopoulos01}
G.~Papadopoulos, P.J. Edwards, and A.F. Murray.
\newblock Confidence estimation methods for neural networks: a practical
  comparison.
\newblock \emph{IEEE Transactions on Neural Networks}, 12:\penalty0 1278, 2001.

\bibitem[Wasserman(2004)]{Wasserman04}
L.~Wasserman.
\newblock \emph{All of Statistics: A Concise Course in Statistical Inference}.
\newblock Springer, 2004.
\newblock ISBN 978-0387402727.

\bibitem[J\'ohannesson et~al.(2002)J\'ohannesson, Bligaard, Ruban, Skriver,
  Jacobsen, and N{\o}rskov]{Johannesson02}
G.H. J\'ohannesson, T.~Bligaard, A.V. Ruban, H.L Skriver, K.W. Jacobsen, and
  J.K. N{\o}rskov.
\newblock Combined electronic structure and evolutionary search approach to
  materials design.
\newblock \emph{Phys. Rev. Lett.}, 88:\penalty0 255506, 2002.

\bibitem[Stucke and Crespi(2003)]{Stucke03}
D.P. Stucke and V.H. Crespi.
\newblock Predictions of new crystalline states for assemblies of
  nanoparticles: Perovskite analogues and 3-d arrays of self-assembled
  nanowires.
\newblock \emph{Nano Letters}, 3:\penalty0 1183, 2003.

\bibitem[Ingber and Rosen(1992)]{Ingber92}
L.~Ingber and B.~Rosen.
\newblock Genetic algorithms and very fast simulated reannealing: A comparison.
\newblock \emph{Mathl. Compat. Modelling}, 16:\penalty0 87, 1992.

\bibitem[Mahfoud and Goldberg(1995)]{Mahfoud95}
S.W. Mahfoud and D.E. Goldberg.
\newblock Parallel recombinative simulated annealing: Agenetic algorithm.
\newblock \emph{Parallel Computing}, 21:\penalty0 1, 1995.

\bibitem[Gelman et~al.(1996)Gelman, Roberts, and Gilks]{Gelman96}
A.~Gelman, G.O. Roberts, and W.R. Gilks.
\newblock Efficient metropolis jumping rules.
\newblock \emph{Bayesian Statistics}, 5:\penalty0 599, 1996.

\bibitem[Halse(2004)]{Halse04}
B.~Halse.
\newblock {Strategic Research Agenda}.
\newblock Technical report, Advisory Council for Aeronautics Research in
  Europe, 2004.

\bibitem[Sims(1984)]{Sims1984}
C.T. Sims.
\newblock A history of superalloy metallurgy for superalloy metallurgists.
\newblock In \emph{Superalloys 1984}, pages 399--419, 1984.
\newblock ISBN 9780895204783.

\bibitem[Reed et~al.(2009{\natexlab{b}})Reed, Tao, and Warnken]{reed2009}
R.~C. Reed, T.~Tao, and N.~Warnken.
\newblock {A}lloys by {D}esign: {A}pplication to {N}ickel-{B}ased {S}ingle
  {C}rystal {S}uperalloys.
\newblock \emph{Acta Materialia}, 57:\penalty0 5898--5913, 2009{\natexlab{b}}.

\bibitem[Connor(2009)]{connor2009}
L.D. Connor.
\newblock \emph{{The development of a dual microstructure heat treated Ni-Base
  superalloy for turbine disc applications}}.
\newblock PhD thesis, University of Cambridge, 2009.

\bibitem[{Special~Metals}(2013)]{specialmetals2013}
{Special~Metals}.
\newblock Special~metals, 2013.
\newblock URL \url{www.specialmetals.com}.

\bibitem[{Haynes~International,~Inc}(2013)]{haynesinternational2013}
{Haynes~International,~Inc}.
\newblock Haynes~international,~inc, 2013.
\newblock URL \url{www.haynesintl.com}.

\bibitem[Tomasello et~al.(1996)Tomasello, Pettit, Birks, Maloney, and
  Radavich]{tomasello1996}
C.M. Tomasello, F.S. Pettit, N.~Birks, J.L. Maloney, and J.F. Radavich.
\newblock Precipitation behaviour in aerex 350.
\newblock In \emph{Superalloys 1996}, pages 145--151. The Minerals, Metals \&
  Materials Society, 1996.
\newblock ISBN 9780873393522.

\bibitem[Radavich and Furrer(2004)]{radavich2004}
J.~Radavich and D.~Furrer.
\newblock Assessment of russian p/m superalloy ep741np.
\newblock In \emph{Superalloys 2004}, pages 381--390. The Minerals, Metals \&
  Materials Society, 2004.
\newblock ISBN 9780873395762.

\bibitem[Sims et~al.(1987)Sims, Stoloff, and Hagel]{sims1987}
C.T. Sims, N.S. Stoloff, and W.C. Hagel.
\newblock \emph{{Superalloys II, High Temperature Materials for Aerospace and
  Industrial Power}}.
\newblock John Wiley \& Sons, Inc, 1987.
\newblock ISBN 9780471011477.

\bibitem[Ewing et~al.(1972)Ewing, Rizzo, and zurLippe]{ewing1976}
B.~Ewing, F.~Rizzo, and C.~zurLippe.
\newblock Powder metallurgy products for advanced gas turbine applications.
\newblock \emph{Superalloys 1972}, pages 1--12, 1972.

\bibitem[Sato and Ohno(1993)]{sato1993}
K.~Sato and T.~Ohno.
\newblock Development of low thermal expansion superalloy.
\newblock \emph{Journal of Materials Engineering and Performance}, 2:\penalty0
  511--516, 1993.

\bibitem[Mannan et~al.(2000)Mannan, Patel, and de~Barbadillo~J.]{mannan2000}
S.~Mannan, S.~Patel, and de~Barbadillo~J.
\newblock Long term thermal stability of inconel alloys 718, 706, 909, and
  waspaloy at 593{C} and 704{C}.
\newblock In \emph{Superalloys 2000}, pages 449--458. The Minerals, Metals \&
  Materials Society, 2000.
\newblock ISBN 9780873394772.

\bibitem[Mitchell(2004)]{mitchell2004}
R.J. Mitchell.
\newblock \emph{{Development of a new powder processed Ni-base superalloy for
  rotor disc application}}.
\newblock PhD thesis, University of Cambridge, 2004.

\bibitem[Meng et~al.(1984)Meng, Sun, Li, and Xie]{meng1984}
Z.~Meng, G.-C. Sun, M.-L. Li, and X.~Xie.
\newblock The strengthening effect of tantalum in nickel-base superalloys.
\newblock \emph{Superalloys 1984}, pages 563--572, 1984.

\bibitem[Tien et~al.(1989)Tien, Collier, and Vignoul]{tien1989}
J.~Tien, J.P. Collier, and G.~Vignoul.
\newblock The role of niobium and other refractory elements in superalloys.
\newblock \emph{Superalloy 718}, pages 553--566, 1989.

\bibitem[Huron et~al.(2004)Huron, Bain, Mourer, Schirra, Reynolds, and
  Montero]{huron2004}
E.S. Huron, K.P. Bain, D.~Mourer, J.~Schirra, P.L. Reynolds, and E.E. Montero.
\newblock The influence of grain boundary elements on properties and
  microstructures of p/m nickel base superalloys.
\newblock In \emph{Superalloys 2004}, pages 73--81. The Minerals, Metals \&
  Materials Society, 2004.
\newblock ISBN 9780873395762.

\bibitem[Sharma and Tewari(1983)]{sharma1983}
K.K. Sharma and S.N. Tewari.
\newblock {A high performance wrought nickel-base superalloy El-929}.
\newblock \emph{Journal of Materials Science}, 18:\penalty0 2915--2922, 1983.

\bibitem[Cowen and Jablonski(2008)]{cowen2008}
C.J. Cowen and P.D Jablonski.
\newblock Elevated temperature mechanical behavior of new low cte superalloys.
\newblock In \emph{Superalloys 2008}, pages 201--208. The Minerals, Metals \&
  Materials Society, 2008.
\newblock ISBN 9780873397285.

\bibitem[Pike(2008)]{pike2008}
L.M. Pike.
\newblock Development of a fabricable gamma prime ($\gamma$') strengthened
  superalloy.
\newblock In \emph{Superalloys 2008}, pages 191--199. The Minerals, Metals \&
  Materials Society, 2008.
\newblock ISBN 9780873397285.

\bibitem[Mannan et~al.(2004)Mannan, Smith, and Patel]{mannan2004}
S.~Mannan, G.D. Smith, and S.~Patel.
\newblock Thermal stability of inconel alloy 783 at 593{C} and 704{C}.
\newblock In \emph{Superalloys 2004}, pages 627--635. The Minerals, Metals \&
  Materials Society, 2004.
\newblock ISBN 9780873395762.

\bibitem[Tillack and Eiselstein(1991)]{tillack1991}
D.J. Tillack and H.L. Eiselstein.
\newblock The invention and definition of alloy 625.
\newblock In \emph{Superalloys 718, 625 and Various Derivatives}, pages 1--14.
  The Minerals, Metals \& Materials Society, 1991.
\newblock ISBN 9780873393522.

\bibitem[Rizzo and Radavich(1991)]{rizzo1991}
F.~J. Rizzo and J.~Radavich.
\newblock Microstructural characterisation of pm 625-type materials.
\newblock In \emph{Superalloys 718, 625 and Various Derivatives}, pages
  297--308. The Minerals, Metals \& Materials Society, 1991.
\newblock ISBN 9780873393522.

\bibitem[Moll et~al.(1971)Moll, Maniar, and Muzyka]{moll1971}
J.H. Moll, G.~N. Maniar, and D.R. Muzyka.
\newblock {Heat treatment of 706 alloy for optimum 1200 F stress rupture
  properties}.
\newblock \emph{Metallurgical and Materials Transactions B}, 2:\penalty0
  2153--2160, 1971.

\bibitem[Brinegar et~al.(1984{\natexlab{a}})Brinegar, Mihalisin, and
  VanderSluis]{brinegar1984}
J.R. Brinegar, J.R. Mihalisin, and J.~VanderSluis.
\newblock The effects of tantalum for columbium substitutions in alloy 713c.
\newblock \emph{Superalloys 1984}, pages 53--61, 1984{\natexlab{a}}.

\bibitem[Bouse and Behrendt(1989)]{bouse1989}
G.K. Bouse and M.R. Behrendt.
\newblock {Mechanical properties of microcast-X alloy 718 fine grain investment
  castings}.
\newblock \emph{Superalloy 718}, pages 319--328, 1989.

\bibitem[Braun and Radavich(1989)]{braun1989}
A.R. Braun and J.~Radavich.
\newblock {A microstructural and mechanical properties comparison of P/M 718
  and P/M TA 718}.
\newblock \emph{Superalloy 718}, pages 623--629, 1989.

\bibitem[Chang and Nahm(1989)]{chang1989}
K.M. Chang and A.H. Nahm.
\newblock {Rene 220: 100 F improvement over alloy 718}.
\newblock \emph{Superalloy 718}, pages 631--646, 1989.

\bibitem[Jackman et~al.(1991)Jackman, Smith, Dix, and Lasonde]{jackman1991}
L.~Jackman, G.~Smith, A.W. Dix, and M.L. Lasonde.
\newblock Rotary forge processing of direct aged inconel 718 for aircraft
  engine shafts.
\newblock In \emph{Superalloys 718, 625 and Various Derivatives}, pages
  125--132. The Minerals, Metals \& Materials Society, 1991.
\newblock ISBN 9780873393522.

\bibitem[Schirra et~al.(1991)Schirra, Caless, and Hatala]{schirra1991}
J.~Schirra, R.H. Caless, and R.W. Hatala.
\newblock The effect of laves phase on the mechanical properties of wrought and
  cast + hip inconel 718.
\newblock In \emph{Superalloys 718, 625 and Various Derivatives}, pages
  375--388. The Minerals, Metals \& Materials Society, 1991.
\newblock ISBN 9780873393522.

\bibitem[Guo et~al.(1991)Guo, Xu, and Loria]{guo1991}
E.~Guo, F.~Xu, and E.A. Loria.
\newblock Comparison of $\gamma$/$\gamma$'' precipitates and mechanical
  properties in modified 718 alloys.
\newblock In \emph{Superalloys 718, 625 and Various Derivatives}, pages
  397--408. The Minerals, Metals \& Materials Society, 1991.
\newblock ISBN 9780873393522.

\bibitem[Mannan et~al.(2003)Mannan, Hibner, and Puckett]{mannan2003}
S.~Mannan, E.~Hibner, and B.~Puckett.
\newblock Physical metallurgy of alloys 718, 725, 725hs, 925 for service in
  aggressive corrosive environments.
\newblock Technical report, Special Metals, 2003.

\bibitem[Rizzo and Buzzanell(1968)]{rizzo1968}
F.~Rizzo and J.D. Buzzanell.
\newblock Effect of chemistry variations on the structural stability of alloy
  718.
\newblock \emph{Superalloys 1968}, pages 501--543, 1968.

\bibitem[Radavich and Meyers(1984)]{radavich1984}
J.~Radavich and D.J. Meyers.
\newblock {Thermomechanical Processing of P/M Alloy 718}.
\newblock \emph{Superalloys 1984}, pages 347--356, 1984.

\bibitem[Xie et~al.(1996)Xie, Liu, Hu, Tang, Xu, Dong, Ni, Zhu, Tien, Zhang,
  and Xie]{xie1996}
X.~Xie, X.~Liu, Y.~Hu, B.~Tang, Z.~Xu, J.~Dong, K.~Ni, Y.~Zhu, S.~Tien,
  L.~Zhang, and W.~Xie.
\newblock The role of phosphorus and sulfur in inconel 718.
\newblock In \emph{Superalloys 1996}, pages 599--606. The Minerals, Metals \&
  Materials Society, 1996.
\newblock ISBN 9780873393522.

\bibitem[Loewenkamp and Radavich(1988)]{loewenkamp1988}
S.A. Loewenkamp and J.F. Radavich.
\newblock {Microstructure and properties of Ni-Fe base Ta-718}.
\newblock \emph{Superalloys 1988}, pages 53--61, 1988.

\bibitem[Quested et~al.(1988)Quested, Mclean, and Winstone]{quested1988}
P.N. Quested, M.~Mclean, and M.R. Winstone.
\newblock {Evaluation of electron-beam, cold hearth refining (EBHCR) of virgin
  and revert IN738LC}.
\newblock \emph{Superalloys 1988}, pages 387--396, 1988.

\bibitem[Seib(2000)]{seib2000}
D.C. Seib.
\newblock Stress rupture behavior of waspaloy an in-738lc at 600c (1112f) in
  low oxygen gaseous environments containing sulfur.
\newblock In \emph{Superalloys 2000}, pages 535--544. The Minerals, Metals \&
  Materials Society, 2000.
\newblock ISBN 9780873394772.

\bibitem[Shaw(1980)]{shaw1980}
S.W.K. Shaw.
\newblock Response of in-939 to process variations.
\newblock In \emph{Superalloys 1980}, pages 275--284. The Minerals, Metals \&
  Materials Society, 1980.
\newblock ISBN 9780871701022.

\bibitem[Sjoeberg et~al.(2004)Sjoeberg, Imamovic, Gabel, Caballero, Brooks,
  Ferte, and Lugan]{sjoeberg2004}
G.~Sjoeberg, D.~Imamovic, J.~Gabel, O.~Caballero, J.W. Brooks, J.P. Ferte, and
  A.~Lugan.
\newblock Evaluation of the in 939 alloy for large aircraft engine structures.
\newblock In \emph{Superalloys 2004}, pages 441--450. The Minerals, Metals \&
  Materials Society, 2004.
\newblock ISBN 9780873395762.

\bibitem[Nganbe and Heilmaier(2009)]{nganbe2009}
M.~Nganbe and M.~Heilmaier.
\newblock {High temperature strength and failure of the Ni-base superalloy
  PM3030}.
\newblock \emph{International Journal of Plasticity}, 25:\penalty0 822--837,
  2009.

\bibitem[Kaufman(1984)]{kaufman1984}
M.~Kaufman.
\newblock Properties of cast mar-m-247 for turbine blisk applications.
\newblock \emph{Superalloys 1984}, pages 43--52, 1984.

\bibitem[Eng and Evans(1980)]{eng1980}
R.D. Eng and D.J. Evans.
\newblock High strength hip consolidated merl 76 disks.
\newblock In \emph{Superalloys 1980}, pages 491--500. The Minerals, Metals \&
  Materials Society, 1980.
\newblock ISBN 9780871701022.

\bibitem[Brinegar et~al.(1984{\natexlab{b}})Brinegar, Norris, and
  Rozenberg]{brinegar1984_2}
J.~Brinegar, L.F. Norris, and L.~Rozenberg.
\newblock Microcast-x fine grain casting - a progress report.
\newblock \emph{Superalloys 1984}, pages 23--32, 1984{\natexlab{b}}.

\bibitem[{Special~Metals}(1971)]{specialmetals1971}
{Special~Metals}.
\newblock Nimonic alloy 263,.
\newblock Technical report, Special~Metals, 1971.
\newblock URL \url{www.specialmetals.com}.

\bibitem[Locq et~al.(2000)Locq, Marty, and Caron]{locq2000}
D.~Locq, M.~Marty, and P.~Caron.
\newblock Optimisation of the mechanical properties of a new pm superalloy for
  disk applications.
\newblock In \emph{Superalloys 2000}, pages 395--403. The Minerals, Metals \&
  Materials Society, 2000.
\newblock ISBN 9780873394772.

\bibitem[Barker and VanDerMolen(1972)]{barker1972}
J.F. Barker and E.H. VanDerMolen.
\newblock {Effect of processing variables on powder metallurgy Rene '95}.
\newblock \emph{Superalloys 1972}, pages 1--23, 1972.

\bibitem[Hunt(2001)]{hunt2001}
D.W. Hunt.
\newblock \emph{The stability \& mechanical properties of a nickel-base turbine
  disc alloy}.
\newblock PhD thesis, Emmanuel College, University of Cambridge, 2001.

\bibitem[Wanner and DeAntonio(1992)]{wanner1992}
E.~Wanner and D.~DeAntonio.
\newblock Development of a new improved controlled thermal expansion superalloy
  with improved oxidation resistance.
\newblock In \emph{Superalloys 1992}, pages 237--246. The Minerals, Metals \&
  Materials Society, 1992.
\newblock ISBN 9780873391894.

\bibitem[Tien et~al.(1990)Tien, Collier, Bretz, and Hendrix]{tien1990}
J.K. Tien, J.~Collier, P.~Bretz, and B.C. Hendrix.
\newblock \emph{High Temperature Materials for Power Engineering}, chapter
  {Raising the high temperature limit of IN718-designing Ticolloy}, pages
  1341--1348.
\newblock Kluwer Academic Publishers, 1990.
\newblock ISBN 0-7923-0925-1.

\bibitem[Gu et~al.(2008)Gu, Cui, Harada, Fukuda, Ping, Mitsuhashi, Kato,
  Kobayashi, and Fujioka]{gu2008}
Y.F. Gu, C.~Cui, H.~Harada, T.~Fukuda, D.~Ping, A.~Mitsuhashi, K.~Kato,
  T.~Kobayashi, and J.~Fujioka.
\newblock Development of ni-co base alloys for high-temperature disk
  applications.
\newblock In \emph{Superalloys 2008}, pages 53--61. The Minerals, Metals \&
  Materials Society, 2008.
\newblock ISBN 9780873397285.

\bibitem[Couturier et~al.(2004)Couturier, Burlet, Terzi, Dubiez, Guetaz, and
  G.]{couturier2004}
R.~Couturier, H.~Burlet, S.~Terzi, S.~Dubiez, L.~Guetaz, and Raisson G.
\newblock Process development and mechanical properties of alloy u720li for
  high temperature turbine disks.
\newblock In \emph{Superalloys 2004}, pages 351--359. The Minerals, Metals \&
  Materials Society, 2004.
\newblock ISBN 9780873395762.

\bibitem[Sczerzenie and Maurer(1984)]{sczerzenie1984}
F.E. Sczerzenie and G.E. Maurer.
\newblock {Development of Udimet 720 for high strength disk applications}.
\newblock \emph{Superalloys 1984}, pages 573--580, 1984.

\bibitem[Jain et~al.(2000)Jain, Ewing, and Yin]{jain2000}
S.K. Jain, B.~Ewing, and C.A. Yin.
\newblock The development of improved performance p/m udimet 720 turbine disks.
\newblock In \emph{Superalloys 2000}, pages 785--794. The Minerals, Metals \&
  Materials Society, 2000.
\newblock ISBN 9780873394772.

\bibitem[Green(1996)]{green1996}
K.A. Green.
\newblock Development of isothermally forged p/m 720 for turbine disk
  applications.
\newblock In \emph{Superalloys 1996}, pages 697--703. The Minerals, Metals \&
  Materials Society, 1996.
\newblock ISBN 9780873393522.

\bibitem[Furrer and Fecht(2000)]{furrer2000}
D.~Furrer and H.J. Fecht.
\newblock Microstructure and mechanical property development in superalloy
  720li.
\newblock In \emph{Superalloys 2000}, pages 415--424. The Minerals, Metals \&
  Materials Society, 2000.
\newblock ISBN 9780873394772.

\bibitem[Ferrari(1976)]{ferrari1976}
A.~Ferrari.
\newblock Effect of microstructure on the early stages of creep deformation of
  an experimental nickel base alloy.
\newblock \emph{Superalloys 1976}, pages 201--213, 1976.

\bibitem[Raisson and Honnorat(1976)]{raisson1976}
G.~Raisson and Y.~Honnorat.
\newblock {P.M. superalloy for high temperature components}.
\newblock \emph{Superalloys 1976}, pages 473--482, 1976.

\bibitem[Richards and Cook(1968)]{richards1968}
E.G. Richards and R.M. Cook.
\newblock Factors influencing the stability of nickel-base high-temperature
  alloys.
\newblock \emph{Superalloys 1968}, pages 1--24, 1968.

\bibitem[Maurer et~al.(1980)Maurer, Jackman, and Domingue]{maurer1980}
G.~Maurer, L.~Jackman, and J.~Domingue.
\newblock Role of cobalt in waspaloy.
\newblock In \emph{Superalloys 1980}, pages 43--52. The Minerals, Metals \&
  Materials Society, 1980.
\newblock ISBN 9780871701022.

\bibitem[Sato et~al.(2010)Sato, Yu-Lung, and Reed]{reed2010}
A.~Sato, C.~Yu-Lung, and R.C. Reed.
\newblock Oxidation of nickel-based single crystal superalloys for industrial
  gas turbine applications.
\newblock \emph{Acta Materialia}, 59:\penalty0 225--240, 2010.

\bibitem[Encinas-Oropesa et~al.(2008)Encinas-Oropesa, Drew, Hardy, Leggett,
  Nicholls, and Simms]{Encinas08}
A.~Encinas-Oropesa, G.L. Drew, M.C. Hardy, A.J. Leggett, J.R. Nicholls, and
  N.J. Simms.
\newblock Effects of oxidation and hot corrosion in a nickel disc alloy.
\newblock In \emph{Superalloys 2008}, pages 609--618. The Minerals, Metals \&
  Materials Society, 2008.
\newblock ISBN 9780873397285.

\bibitem[Gayda et~al.(2002)Gayda, Kantzos, and Miller]{Gayda02}
J.~Gayda, P.~Kantzos, and J.~Miller.
\newblock Quench crack behavior of nickel-base disk superalloys.
\newblock \emph{NASA/TM}, 2002.

\bibitem[Conduit et~al.(2014{\natexlab{b}})Conduit, Conduit, Stone, and
  Hardy]{Conduit2013ii}
B.D. Conduit, G.J. Conduit, H.J. Stone, and M.C. Hardy.
\newblock Development of a new nickel based superalloy for a combustor liner
  and other high temperature applications.
\newblock Patent GB1408536, 2014{\natexlab{b}}.

\bibitem[Conduit et~al.(2014{\natexlab{c}})Conduit, Conduit, Stone, and
  Hardy]{Conduit2014v}
B.D. Conduit, G.J. Conduit, H.J. Stone, and M.C. Hardy.
\newblock Molybdenum-hafnium alloys for high temperature applications.
\newblock Patent EP14161255, US 2014/223465, 2014{\natexlab{c}}.

\bibitem[Conduit et~al.(2014{\natexlab{d}})Conduit, Conduit, Stone, and
  Hardy]{Conduit2014vi}
B.D. Conduit, G.J. Conduit, H.J. Stone, and M.C. Hardy.
\newblock Molybdenum-niobium alloys for high temperature applications.
\newblock Patent EP14161529, US 2014/224885, 2014{\natexlab{d}}.

\bibitem[{American Elements}(2013)]{AmericanElements2013}
{American Elements}.
\newblock Element catalogue, 2013.
\newblock URL \url{https://www.americanelements.com}.

\end{thebibliography}

\end{document}